\documentclass[12pt]{iopart}
\usepackage{graphicx}

\usepackage{amssymb}
\usepackage{longtable}
\usepackage{amsthm}
\usepackage{cite}

\usepackage{bm}

\begin{document}
\jpb

\title
{Theoretical investigation of spectroscopic properties of W$^{26+}$ 
in EBIT plasma}

\author{
V. Jonauskas, A. Kynien\.{e}, P. Rynkun, S. Ku\v{c}as, G. Gaigalas, 
R. Kisielius, \v{S}. Masys, G. Merkelis, L. Rad\v{z}i{\=u}t\.e
}

\address{
Institute of Theoretical Physics and Astronomy, Vilnius
University, A. Go\v{s}tauto 12,  Vilnius, LT-01108, Lithuania
}

\ead{Valdas.Jonauskas@tfai.vu.lt}

\begin{abstract}
Energy levels, radiative transition wavelengths and probabilities have been 
studied for the W$^{26+}$ ion using multiconfiguration Dirac-Fock and 
Dirac-Fock-Slater methods. Corona and collisional-radiative models have been 
applied to determine lines and corresponding configurations in a low-density 
electron beam ion trap (EBIT) plasma. Correlation effects for the $4f^{2}$,  
$4d^{9}4f^{3}$, $4f5l$ ($l=0,...,4$), $4fng$ ($n=5, 6, 7$) configurations have 
been estimated by presenting configuration interaction strengths. It was 
determined that correlation effects are important for the 
$4f5s \rightarrow 4f^{2}$ transitions corresponding to weak electric octupole 
transitions in a single-configuration approach. Correlation effects influence 
the $4f5d \rightarrow 4f^{2}$ transitions by increasing transition probabilities 
by an order of magnitude. Identification of some lines observed in fusion plasma
has been proposed.  Spectra modeling shows strong increase of lines 
originating from the $4f5s \rightarrow 4f^{2}$ transitions. Other transitions 
from the $10-30$ nm region can be of interest for the EBIT plasma.

\end{abstract}



\maketitle

\section{Introduction}

Tungsten is a primary candidate as a plasma-facing material in fusion devices 
due to many important properties. Unfortunately, emission of the tungsten ions 
that penetrate central regions of the fusion plasma leads to undesirable energy 
losses. The concentration of these ions has to be monitored in order to create 
and maintain the fusion reaction. Thus, reliable atomic data for various 
tungsten ions are needed for successful control of processes in the fusion 
plasma.  

The most intense tungsten emission in the fusion plasma occurs at around 
5 nm where a quasicontinuum band is formed \cite{1977pla_63_295_isler, 
1978pla_66_109_hinnov, 1988pla_127_255_Finkenthal, 2007pfr_2_s1060_chowdhuri, 
2011cjp_89_591_podpaly, 2008ppcp_50_085016_Putterich, 2010jpb_43_0953_harte}.   
Collisional-radiative modeling (CRM) showed large contributions to the spectral 
region from the W$^{27+}$ -- W$^{37+}$ ions at about $2$ keV electron 
temperature \cite{2008ppcp_50_085016_Putterich}. It was also predicted that the 
lower ionization stages down to W$^{21+}$ strongly contribute to the emission 
at about $1$ keV electron temperature  \cite{2008ppcp_50_085016_Putterich}. 
Additional structure of lines at 6 nm with lower intensity than the main peak 
at 5 nm is observed in the fusion spectra. A careful examination of these 
lines predicts that they are formed by ionization stages in the range between 
W$^{21+}$ and W$^{35+}$  \cite{2008ppcp_50_085016_Putterich}.
The modeling for tungsten emission between 
10 and 30 nm indicated contributions from the lower than W$^{28+}$ charged 
states \cite{2008ppcp_50_085016_Putterich}. The region around 20 nm has been 
studied in a fusion plasma  of the Large Helical Device (LHD)
\cite{2011jpb_44_175004_suzuki}. It was also found  
that the stages lower than W$^{27+}$ are the main contributors to the 
emission spectrum. Furthermore, large 
contributions from the $6g \rightarrow 4f$ and $5g \rightarrow 4f$ transitions 
in the W$^{24+}$ to W$^{27+}$ ions have been observed at the 
$1.5 - 3.5$\,nm region using Compact electron Beam Ion Trap (CoBIT) and LHD 
\cite{2012aipcp_1438_91_sakaue, 2013aipcp_1545_143_morita}. Modeled spectra of
W$^{23+}$  have illustrated the importance of ions with open $f$ shells in 
the formation of the fusion spectra \cite{2013acp_1545_132_putterich}.

The ground configurations of the W$^{15+}$ -- W$^{27+}$ ions have open $f$ 
shells. Ions with the open $f$ shells have been rarely studied theoretically 
due to complexity of calculations. Configurations of such ions possess a large 
number of energy levels. Many energy levels of different configurations overlap,
indicating importance of correlation effects. Furthermore, investigations of 
high-$Z$ elements require relativistic effects to be considered in the 
Dirac-Fock approach with quantum electrodynamic (QED) corrections. 

Transitions from many ions contribute to the line-of-sight measurements in the 
fusion plasma. The electron beam ion trap (EBIT) devices provide a unique 
opportunity to study emission mainly from the desirable ionization stage 
determined by the energy of the electron beam. Due to this feature, populations 
of neighboring ionization states are less expressed. Furthermore, such plasma 
features low density of electrons. That  leads to the dominant population of the 
ground and long-lived levels. The emission around $5$ nm has been observed in 
the EBIT plasma for the W$^{21+}$ -- W$^{46+}$ ions 
\cite{2001pra_64_012720_radtke, 2007apmidf_13_45_radtke}. High ionization stages
 of tungsten have been studied using the NIST EBIT device and 
collisional-radiative modeling \cite{2006pra_74_042514_ralchenko, 
2007jpb_40_3861_ralchenko, 2011pra_83_032517_ralchenko}. The CoBIT was used to 
analyze emission spectra from the W$^{23+}$ to W$^{33+}$ ions in the $1.5 - 3.5$
nm range \cite{2012aipcp_1438_91_sakaue, 2013aipcp_1545_143_morita}.

The main aim of the current work is to determine the strongest lines in the 
spectrum of the W$^{26+}$ ion by performing the corona and 
collisional-radiative modeling of spectral lines and to estimate influence 
of correlation effects for the configurations corresponding to the strongest 
lines. Previous theoretical investigations of spectra from the W$^{26+}$ ion 
have been performed using pseudorelativistic approach 
\cite{2008ppcp_50_085016_Putterich, 2010jpb_43_0953_harte, 
2011jpb_44_175004_suzuki, 2012jpb_45_205002_harte}.  Our work considers 
modeling for a monoenergetic electron beam that corresponds to the EBIT 
measurements. HULLAC code \cite{BarShalomHULLAC} has been used previously 
to model the W$^{26+}$ ion spectrum in the EBIT plasma by applying a 
collisional-radiative approach \cite{2007apmidf_13_45_radtke}.
However, that study included only 461 levels, 
and only the wavelengths and intensities of the strongest lines were presented 
in the vicinity of $5$ nm. Furthermore, this modeling did not involve the 
$4f5g$ configuration to which strong excitations  from the ground configuration
occur.  Energy levels of the ground configuration and the magnetic dipole and 
electric quadrupole transitions among these levels have also been studied using 
the multiconfiguration Dirac-Fock (MCDF) method \cite{2011jpb_44_145004_ding}. 
Recently, an extended investigation of the energy levels of the ground 
configuration of the W$^{26+}$ ion has been presented using MCDF and the 
multireference relativistic many-body perturbation theory (MR-RMBPT) 
calculations \cite{2014pra_90_052517_fei}. Studies of CoBIT and LHD spectra 
predicted contributions from the $6g \rightarrow 4f$ and $5g \rightarrow 4f$ 
transitions \cite{2012aipcp_1438_91_sakaue, 2013aipcp_1545_143_morita}. 

Investigations of the W$^{29+}$ -- W$^{37+}$ ions in the EBIT plasma 
demonstrated that relative line intensities calculated by studying excitations 
from the corresponding ground levels are in quite good agreement with the data 
from the collisional-radiative modeling \cite{2007jpb_40_2179_jonauskas}.  
There the electron-impact excitation rates were considered as being 
proportional to the electric dipole transition probabilities because the 
plane-wave Born  matrix element transforms to  the matrix elements of the 
electric multipole transition operators with additional factors. Therefore, the
corresponding selection rules of the electric multipole transitions are 
applicable to the plane-wave Born transitions. The use of the plane-wave Born 
approximation is justified when the incident electron energies are much greater 
than the excitation ones. 

Therefore, two approaches are used for corona modeling spectral lines of the 
W$^{26+}$ ion in this work. In the first one, electron-impact excitation rates 
are determined in the distorted wave (DW) approximation. In the second approach, 
the electric dipole line strengths are used instead of the electron-impact 
excitation rates because the leading term of the first order is proportional to 
the electric dipole (spin-allowed) transition probability divided by the third 
power of transition energy for the collision cross sections within the plane-wave 
Born approximation, i.e., is proportional to the transition line strength.
On one hand, in the pure $LS$ coupling, this approach applies to the spin-allowed 
transitions. On other hand, since we use intermediate coupling CI wavefunctions,
selection rules for the total-spin quantum number do not apply, and only the
total angular momentum $J$ remains a valid description of the fine-structure 
level. The plane-wave Born approximation neglects the excitations, which
correspond to other than E1 radiative transition types, such as the magnetic 
dipole or electric quadrupole transitions. But in this case we must point out
that the excitations corresponding to the M1 and E2 radiative transitions go
to the configurations of the same as the ground configuration parity. 
When the ground configuration fine-structure levels are excited, resulting
radiative decay transitions form the lines that have wavelengths beyond the 
scope of this work. These lines have already been studied theoretically and 
experimentally \cite{2011jpb_44_145004_ding, 2014pra_90_052517_fei}. 
Furthermore, the levels populated by the neglected excitations to the higher 
even-parity configurations decay by the radiative E1 cascades down to the 
intermediate states of the odd-parity configurations that have strong decay 
channels to the ground configuration. These cascades are included in our 
calculations and have a large impact on the populations of levels. 
In the current work, we investigate the excitations from all ground 
configuration levels (rather that the lowest one) with subsequent radiative 
cascades.

In addition, a comparison with the CRM is used to demonstrate that simplified 
approaches are applicable even for such complex systems when the EBIT spectra
are analyzed.

In the next section we present the corona and collisional-radiative models and 
the MCDF method used to calculate the energy levels and radiative transition 
probabilities. In Sec. 3, the obtained results for the energy levels and 
emission spectra are discussed, and, in Sec. 4, spectra from the corona model 
are presented and discussed.

\section{Method of calculation}

The corona modeling of spectral lines has been performed for excitations from 
the levels of the ground configuration of the W$^{26+}$ ion. Population of the 
excited levels for the excitation from the level is expressed as 
\begin{equation}
n_{i} = \frac{N_{e} C_{mi}}{\sum_{k<i} A_{ik}},
\label{corona}
\end{equation}
where $m$ is the index of the level from which the excitations are studied;  
$C_{mi}$ - the electron impact excitation/deexcitation rate from the level $m$ 
to the level $i$, $A_{ik}$ - a radiative transition probability.

Population of levels by radiative cascades from the higher-lying levels is taken 
into account by the following expression:
\begin{equation}
\frac{\sum_{m>i} n_{m} A_{mi}}{\sum_{k<i} A_{ik}}.
\label{cascade}
\end{equation}

Initial populations of the levels of the ground configuration are assumed to 
be equal to their statistical weights. The spectra, originating following 
excitations from these levels, are summed resulting in the total emission 
spectrum. In addition, the CRM is used to check the accuracy of the applied 
corona models. Populations of levels in the CRM have been determined by solving 
the system of coupled rate equations:

\begin{equation}
\frac{d n_{i}(t)}{dt} = N_{e} \sum_{k} n_{k}(t) C_{ki} +  \sum_{k>i} n_{k}(t) A_{ki} - N_{e} n_{i}(t) \sum_{k} C_{ik} - n_{i}(t) \sum_{j<i} A_{ij}
\label{crm}
\end{equation} 
in the steady-state equilibrium approximation ($\frac{d n_{i}}{dt} = 0$). Here 
$n_{i}$ is the population of the level $i$, $N_{e}$ is the electron density 
($N_{e} = 1 \times 10^{12}$ cm$^{-3}$, which was the approximate electron 
density in the EBIT measurements \cite{2001pra_64_012720_radtke}.).

The GRASP2K code \cite{2013cpc_184_9_jonsson} is used to calculate the 
wavefunctions as well as the matrix elements of the Dirac-Coulomb-Breit 
Hamiltonian and the radiative transition operators (electric dipole, quadrupole, 
octupole as well as magnetic dipole and quadrupole). The Dirac-Coulomb-Breit 
Hamiltonian consists of the one-electron Dirac Hamiltonian, the Coulomb 
repulsion operator, and the transverse interaction operator 
(which corresponds to the Breit interaction in the low-frequency limit). 

The QED corrections are considered in the first-order perturbation theory. They 
include the vacuum polarization and the self-energy (known as the Lamb shift). 
The correlation corrections are taken into account by the relativistic 
configuration interaction (RCI) method. Finally, the effects of the finite 
nuclear size are modeled by using a two-component Fermi statistical distribution 
function.

In addition to the GRASP2K \cite{2013cpc_184_9_jonsson} calculations, the 
Flexible Atomic Code (FAC) \cite{2008cjp_86_675_Gu}, which incorporates the 
Dirac-Fock-Slater method, is employed to obtain energy levels, radiative 
transition probabilities, and electron-impact excitation rates in the DW 
approach. The same basis of configurations is used in both calculations. 
The electron-impact excitation rates from the levels of the ground 
configuration are calculated at the electron beam energy of $833$ eV and the 
electron beam density of $10^{12}$ cm$^{-3}$. The Gaussian distribution 
function with a full width at half-maximum of $30$ eV is used for the electron 
energy. Thus, we can estimate accuracy of the corona modeling with the MCDF 
data when the electric dipole line strengths are used instead of electron-impact
excitation rates at various wavelengths. On the other hand, atomic data produced
by different codes for the W$^{26+}$ ion can be compared.

Configuration interaction strength (CIS) is used to estimate the configuration 
interaction between two configurations $K_{1}$ and $K_{2}$ 
\cite{Karazija1, 1997ps_55_667_kucas}:
\begin{equation}
T(K_{1},K_{2})=\frac{\sum\limits_{\gamma_{1} \gamma_{2}} \langle 
\Phi(K_{1}\gamma_{1}) \vert H_{ \mathrm{DC}} \vert \Phi(K_{2}\gamma_{2})
\rangle ^2}{\bar{E}(K_{1},K_{2}) ^2},
\label{ccis}
\end{equation}
where the quantity in the numerator is the interconfiguration matrix element of
the Dirac-Coulomb Hamiltonian $H_{ \mathrm{DC}}$ and $\bar{E}(K_{1},K_{2})$ is 
the mean energy distance between the interacting levels of configurations 
$K_{1}$ and $K_{2}$:
\begin{eqnarray}
\hspace{-2.5cm}
\bar{E}(K_{1},K_{2}) =
\nonumber \\
&& \hspace{-2.5cm}
=\frac{\sum\limits_{\gamma_{1} \gamma_{2}} \left[ \langle \Phi(K_{1}\gamma_{1}) \vert H_{ \mathrm{DC}}
\vert \Phi(K_{1}\gamma_{1})\rangle - \langle \Phi(K_{2}\gamma_{2}) \vert H_{ \mathrm{DC}} \vert\Phi( K_{2}\gamma_{2})\rangle \right]
\langle \Phi(K_{1}\gamma_{1}) \vert H_{ \mathrm{DC}} \vert \Phi(K_{2}\gamma_{2})\rangle ^2}
{\sum\limits_{\gamma_{1} \gamma_{2}} \langle \Phi(K_{1} \gamma_{1}) \vert H_{\mathrm{DC}} \vert \Phi(K_{2} \gamma_{2})\rangle ^2}. 
\nonumber \\
\label{ekk} 
\end{eqnarray}
The summation in (\ref{ccis}) and (\ref{ekk})  is performed over all states 
$\gamma_{1}$ and $\gamma_{2}$ of the  configurations $K_{1}$ and $K_{2}$, 
respectively.

The $T(K_{1},K_{2})$ value, divided by the statistical weight $g_{1}=g(K_{1})$ 
of the analyzed configuration $K_{1}$ has the meaning of the average weight of 
the admixed configuration $K_{2}$ in the expansion of the wave function for 
$K_{1}$. CIS has been successfully applied for the investigation of energy 
levels \cite{2015jqsrt_152_94_radziute}, Auger cascades 
\cite{2008jpb_41_215005_jonauskas, 2010pra_82_043419_palaudoux, 
2011pra_84_053415_jonauskas}, electric dipole \cite{2007ljp_47_249_kucas} and 
magnetic dipole \cite{2010pra_81_012506_jonauskas, 2012adndt_98_19_jonauskas} 
transitions.

\section{Energy levels and radiative transition probabilities}
\label{sec:gA}


\begin{figure}
 \includegraphics[scale=0.5]{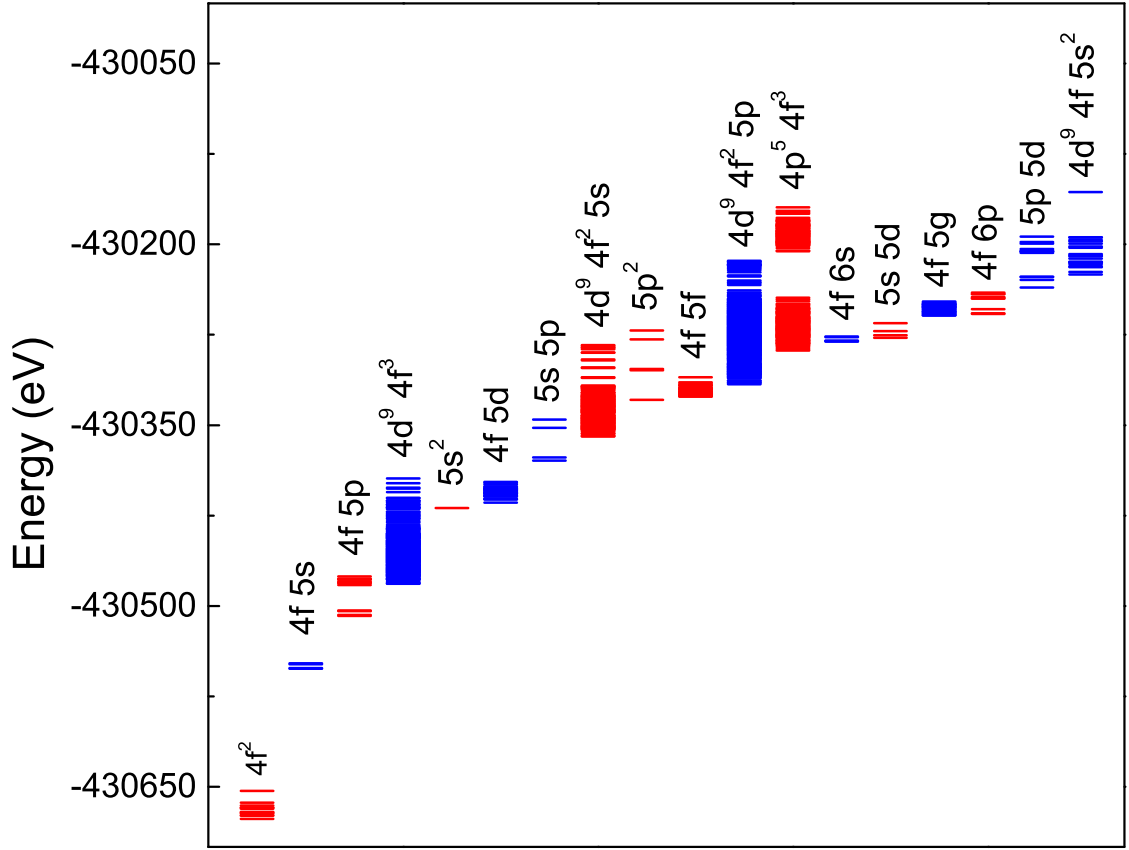}
 \caption{
\label{w26energy} 
Energy levels of the lowest configurations in W$^{26+}$. 
}
\end{figure}


Energy levels of the lowest configurations of W$^{26+}$ are presented in 
Fig.~\ref{w26energy}. The total number of configurations included in the 
present study amounts to 44, they produce 11594 levels. These configurations 
correspond to one-electron promotions from the $4f^{2}$ and $4f5s$ 
configurations. In addition, the radiative transition wavelengths and probabilities have been 
studied for some transitions using a selected basis of configurations in order 
to investigate the importance of correlation effects. 

The ground configuration of the W$^{26+}$ ion consists of 13 levels. 
Table~\ref{energy} presents levels of the W$^{26+}$ ion with the total radiative 
decay rates smaller than the ones of the ground configuration levels. It can be 
seen that FAC energy levels are slightly higher than GRASP2K energy levels.
The largest radiative transition probability of the ground configuration levels 
corresponds to the level 13 with $J=0$ (see Table~\ref{lft}). It is the highest 
level of the ground configuration. It can be seen from Table~\ref{lft} that 
there are many levels of the $4d^{9}4f^{3}$ configuration with the radiative 
lifetimes greater than $2.838 \times 10^{-4}$ s$^{-1}$. These levels have 
extremely large $J$ values, and the radiative decay paths from them are limited 
by selection rules for radiative transitions. Good agreement among GRASP2K and 
FAC transition probabilities is observed.

\begingroup

\renewcommand{\arraystretch}{1.3}
\renewcommand{\tabcolsep}{1mm}

\scriptsize

\begin{flushleft}

\setcounter{table}{0}
\LTcapwidth 15cm
\begin{longtable}{@{}r r r r l}
\caption{\label{energy}
GRASP2K and FAC calculated energy levels relative to the ground energy of the ion {W}$^{26+}$ 
with spectroscopic identifications. Levels having the largest lifetimes and levels 
to which radiative transition takes place from these levels are presented. 
$J$ and P stands for the total angular momentum quantum number $J$ and the parity P.  
$E_{\mathrm{ground}}$ = $-15827.06957$ a.u.(GRASP) = $-15826.17202$ a.u. (FAC).}\\
\hline
No &     $J$P & $E_{\mathrm{GRASP}}$ (a.u.) & $E_{\mathrm{FAC}}$ (a.u.)  &   Composition \\ 
\hline
\endfirsthead

\caption[]{(continued)}  \\
\hline
 No &     $J$P & $E_{\mathrm{GRASP}}$ (a.u.) & $E_{\mathrm{FAC}}$ (a.u.) &    Composition \\ 
\hline
\endhead
\hline \multicolumn{5}{r}{\textit{Continued on next page}} \\
\endfoot
\hline
\endlastfoot

     1 &   4$+$   &  $E_{\mathrm{ground}}$ & $E_{\mathrm{ground}}$ &     88\%  4f$^2$ ($_1^3$H) $^3$H
                                     $+$  11\%  4f$^2$ ($_1^1$G) $^1$G
                                     $+$   1\%  4f$^2$ ($_1^3$F) $^3$F  \\
     2 &   2$+$   &      0.08906  & 0.08917 &     87\%  4f$^2$ ($_1^3$F) $^3$F
                                     $+$  12\%  4f$^2$ ($_1^1$D) $^1$D
                                     $+$   1\%  4f$^2$ ($_1^3$P) $^3$P  \\
     3 &   5$+$   &      0.11418  & 0.11235 &    100\%  4f$^2$ ($_1^3$H) $^3$H  \\
     4 &   4$+$   &      0.17475  & 0.17312 &     47\%  4f$^2$ ($_1^1$G) $^1$G
                                     $+$  44\%  4f$^2$ ($_1^3$F) $^3$F
                                     $+$   9\%  4f$^2$ ($_1^3$H) $^3$H  \\
     5 &   3$+$   &      0.17597  & 0.17451 &     99\%  4f$^2$ ($_1^3$F) $^3$F  \\
     6 &   6$+$   &      0.21176  & 0.20862 &    94\%  4f$^2$ ($_1^3$H) $^3$H
                                     $+$   6\%  4f$^2$ ($_1^1$I) $^1$I  \\
     7 &   4$+$   &      0.30939  & 0.30574 &     55\%  4f$^2$ ($_1^3$F) $^3$F
                                     $+$  42\%  4f$^2$ ($_1^1$G) $^1$G
                                     $+$   2\%  4f$^2$ ($_1^3$H) $^3$H  \\
     8 &   2$+$   &      0.33104  & 0.33047 &     57\%  4f$^2$ ($_1^1$D) $^1$D
                                     $+$  32\%  4f$^2$ ($_1^3$P) $^3$P
                                     $+$  10\%  4f$^2$ ($_1^3$F) $^3$F  \\
     9 &   0$+$   &      0.34915  & 0.34979 &     93\%  4f$^2$ ($_1^3$P) $^3$P
                                     $+$   7\%  4f$^2$ ($_1^1$S) $^1$S  \\
    10 &   1$+$   &      0.40082  & 0.40089 &     99\%  4f$^2$ ($_1^3$P) $^3$P  \\
    11 &   6$+$   &      0.40702  & 0.40591 &     94\%  4f$^2$ ($_1^1$I) $^1$I
                                     $+$   6\%  4f$^2$ ($_1^3$H) $^3$H  \\
    12 &   2$+$   &      0.48727  & 0.48505 &     67\%  4f$^2$ ($_1^3$P) $^3$P
                                     $+$  31\%  4f$^2$ ($_1^1$D) $^1$D
                                     $+$   2\%  4f$^2$ ($_1^3$F) $^3$F  \\
    13 &   0$+$   &      0.84969  & 0.84986 &    93\%  4f$^2$ ($_1^1$S) $^1$S
                                     $+$   7\%  4f$^2$ ($_1^3$P) $^3$P  \\
    31 &   6$-$   &      7.28864  & 7.32784 &     34\%  4d$^9$ 4f$^3$ ($_1^4$I) $^5$H
                                     $+$  24\%  4d$^9$ 4f$^3$ ($_1^4$I) $^5$I
                                     $+$  13\%  4d$^9$ 4f$^3$ ($_1^4$G) $^5$H  $+$ \\
       &          &                 &  &  8\%  4d$^9$ 4f$^3$ ($_1^4$I) $^5$K
                                     $+$   5\%  4d$^9$ 4f$^3$ ($_1^4$G) $^5$I  \\
    40 &   7$-$   &      7.33121  & 7.36437 &     36\%  4d$^9$ 4f$^3$ ($_1^4$I) $^5$L
                                     $+$  23\%  4d$^9$ 4f$^3$ ($_1^4$I) $^3$L
                                     $+$  19\%  4d$^9$ 4f$^3$ ($_1^4$I) $^5$K  $+$ \\
       &          &                 &  &    6\%  4d$^9$ 4f$^3$ ($_2^2$H) $^3$K
                                     $+$   5\%  4d$^9$ 4f$^3$ ($_1^4$I) $^5$I  \\
    44 &   8$-$   &      7.37013  & 7.40683 &     58\%  4d$^9$ 4f$^3$ ($_1^4$I) $^5$L
                                     $+$  18\%  4d$^9$ 4f$^3$ ($_1^4$I) $^5$K
                                     $+$  16\%  4d$^9$ 4f$^3$ ($_1^4$I) $^3$L  $+$ \\
       &          &                 &    &  3\%  4d$^9$ 4f$^3$ ($_2^2$H) $^3$K
                                     $+$   2\%  4d$^9$ 4f$^3$ ($_1^4$I) $^5$I  \\
    45 &   7$-$   &      7.37264  & 7.41288 &     27\%  4d$^9$ 4f$^3$ ($_1^4$I) $^5$H
                                     $+$  24\%  4d$^9$ 4f$^3$ ($_1^4$I) $^5$I
                                     $+$  17\%  4d$^9$ 4f$^3$ ($_1^4$G) $^5$H  $+$ \\
       &          &                 &    &  8\%  4d$^9$ 4f$^3$ ($_1^4$G) $^5$I
                                     $+$   7\%  4d$^9$ 4f$^3$ ($_1^4$I) $^3$L  \\
    51 &   9$-$   &      7.40793  & 7.44710 &     73\%  4d$^9$ 4f$^3$ ($_1^4$I) $^5$L
                                     $+$  11\%  4d$^9$ 4f$^3$ ($_1^4$I) $^5$K
                                     $+$   7\%  4d$^9$ 4f$^3$ ($_1^4$I) $^3$L  $+$ \\
       &          &                 &   &   4\%  4d$^9$ 4f$^3$ ($_1^2$K) $^3$M
                                     $+$   2\%  4d$^9$ 4f$^3$ ($_1^2$K) $^1$M  \\
    54 &   7$-$   &      7.42581  & 7.46300 &     26\%  4d$^9$ 4f$^3$ ($_1^4$I) $^5$K
                                     $+$  15\%  4d$^9$ 4f$^3$ ($_1^4$I) $^3$L
                                     $+$  11\%  4d$^9$ 4f$^3$ ($_1^4$G) $^5$H   $+$ \\
       &          &                 &   &  8\%  4d$^9$ 4f$^3$ ($_1^4$I) $^5$H
                                     $+$   7\%  4d$^9$ 4f$^3$ ($_1^4$I) $^3$I  \\
    57 &  10$-$   &      7.44170  & 7.48247 &     79\%  4d$^9$ 4f$^3$ ($_1^4$I) $^5$L
                                     $+$  19\%  4d$^9$ 4f$^3$ ($_1^2$K) $^3$M
                                     $+$   1\%  4d$^9$ 4f$^3$ ($_1^2$L) $^3$N  $+$ \\ 
       &          &                 &    &  1\%  4d$^9$ 4f$^3$ ($_1^2$L) $^1$N  \\
    62 &   8$-$   &      7.47234  & 7.51131 &     33\%  4d$^9$ 4f$^3$ ($_1^4$I) $^5$K
                                     $+$  27\%  4d$^9$ 4f$^3$ ($_1^4$I) $^5$I
                                     $+$  14\%  4d$^9$ 4f$^3$ ($_1^4$I) $^3$L  $+$ \\
       &          &                 &    & 13\%  4d$^9$ 4f$^3$ ($_1^4$G) $^5$I
                                     $+$   4\%  4d$^9$ 4f$^3$ ($_2^2$H) $^3$K  \\
    68 &   7$-$   &      7.52623  & 7.56236 &     21\%  4d$^9$ 4f$^3$ ($_1^4$F) $^5$H
                                     $+$  19\%  4d$^9$ 4f$^3$ ($_1^4$I) $^3$L
                                     $+$  10\%  4d$^9$ 4f$^3$ ($_1^2$G) $^3$I  $+$ \\
       &          &                 &    &  8\%  4d$^9$ 4f$^3$ ($_1^4$I) $^3$K
                                     $+$   7\%  4d$^9$ 4f$^3$ ($_2^2$G) $^3$I  \\
    75 &   9$-$   &      7.55210  & 7.59231 &     70\%  4d$^9$ 4f$^3$ ($_1^4$I) $^5$K
                                     $+$  18\%  4d$^9$ 4f$^3$ ($_1^4$I) $^3$L
                                     $+$   5\%  4d$^9$ 4f$^3$ ($_1^2$K) $^3$L  $+$ \\
       &          &                 &     & 5\%  4d$^9$ 4f$^3$ ($_1^4$I) $^5$L
                                     $+$   1\%  4d$^9$ 4f$^3$ ($_1^2$K) $^1$M  \\
    76 &   8$-$   &      7.55504  & 7.59386 &     22\%  4d$^9$ 4f$^3$ ($_1^4$I) $^5$K
                                     $+$  19\%  4d$^9$ 4f$^3$ ($_1^4$G) $^5$I
                                     $+$  15\%  4d$^9$ 4f$^3$ ($_1^4$I) $^3$L  $+$ \\
       &          &                 &    & 13\%  4d$^9$ 4f$^3$ ($_1^4$I) $^3$K
                                     $+$  11\%  4d$^9$ 4f$^3$ ($_1^4$I) $^5$I  \\
    82 &   9$-$   &      7.57725  & 7.61564 &     51\%  4d$^9$ 4f$^3$ ($_1^2$K) $^3$M
                                     $+$  26\%  4d$^9$ 4f$^3$ ($_1^2$K) $^1$M
                                     $+$   9\%  4d$^9$ 4f$^3$ ($_1^2$I) $^3$L  $+$ \\
       &          &                 &   &   6\%  4d$^9$ 4f$^3$ ($_1^4$I) $^5$L
                                     $+$   4\%  4d$^9$ 4f$^3$ ($_1^2$K) $^3$L  \\
    85 &  10$-$   &      7.58743  & 7.62760 &     39\%  4d$^9$ 4f$^3$ ($_1^2$K) $^3$M
                                     $+$  26\%  4d$^9$ 4f$^3$ ($_1^2$L) $^3$N
                                     $+$  17\%  4d$^9$ 4f$^3$ ($_1^4$I) $^5$L  $+$ \\ 
       &          &                 &   &  17\%  4d$^9$ 4f$^3$ ($_1^2$L) $^1$N
                                     $+$   1\%  4d$^9$ 4f$^3$ ($_1^2$L) $^3$M  \\
    95 &   8$-$   &      7.63149  &  7.67051 &     30\%  4d$^9$ 4f$^3$ ($_1^2$H) $^3$K
                                     $+$  16\%  4d$^9$ 4f$^3$ ($_1^2$I) $^3$K
                                     $+$  16\%  4d$^9$ 4f$^3$ ($_1^2$I) $^3$L  $+$ \\
       &          &                 &          &12\%  4d$^9$ 4f$^3$ ($_1^2$K) $^3$L
                                     $+$   5\%  4d$^9$ 4f$^3$ ($_1^2$K) $^3$M  \\
   112 &  11$-$   &      7.69139  & 7.73386 &    100\%  4d$^9$ 4f$^3$ ($_1^2$L) $^3$N  \\
   117 &   9$-$   &      7.71709  & 7.75823 &     56\%  4d$^9$ 4f$^3$ ($_1^2$I) $^3$L
                                     $+$  15\%  4d$^9$ 4f$^3$ ($_1^2$K) $^1$M
                                     $+$  13\%  4d$^9$ 4f$^3$ ($_1^2$K) $^3$L $+$ \\ 
       &          &                 &   &   7\%  4d$^9$ 4f$^3$ ($_1^2$L) $^3$L
                                     $+$   4\%  4d$^9$ 4f$^3$ ($_1^2$K) $^3$M  \\
   121 &  10$-$   &      7.72446  & 7.76756 &    42\%  4d$^9$ 4f$^3$ ($_1^2$K) $^3$M
                                     $+$  30\%  4d$^9$ 4f$^3$ ($_1^2$L) $^3$N
                                     $+$  24\%  4d$^9$ 4f$^3$ ($_1^2$L) $^1$N  $+$ \\
       &          &                 &  &    4\%  4d$^9$ 4f$^3$ ($_1^4$I) $^5$L  \\
   127 &   8$-$   &      7.74347  & 7.78090 &     25\%  4d$^9$ 4f$^3$ ($_1^2$K) $^3$M
                                     $+$  24\%  4d$^9$ 4f$^3$ ($_1^2$K) $^1$L
                                     $+$  21\%  4d$^9$ 4f$^3$ ($_1^2$K) $^3$L  $+$ \\
       &          &                 &   &   9\%  4d$^9$ 4f$^3$ ($_1^2$H) $^3$K
                                     $+$   5\%  4d$^9$ 4f$^3$ ($_1^2$I) $^1$L  \\
   133 &   8$-$   &      7.76413  & 7.80236 &     45\%  4d$^9$ 4f$^3$ ($_1^4$G) $^5$I
                                     $+$  10\%  4d$^9$ 4f$^3$ ($_1^4$I) $^5$I
                                     $+$   8\%  4d$^9$ 4f$^3$ ($_1^2$K) $^3$K  $+$ \\
       &          &                 &   &   8\%  4d$^9$ 4f$^3$ ($_1^4$I) $^3$K
                                     $+$   6\%  4d$^9$ 4f$^3$ ($_1^2$I) $^3$K  \\
   152 &   9$-$   &      7.82012  & 7.85768 &    31\%  4d$^9$ 4f$^3$ ($_1^2$K) $^3$L
                                     $+$  24\%  4d$^9$ 4f$^3$ ($_1^2$I) $^3$L
                                     $+$  14\%  4d$^9$ 4f$^3$ ($_1^2$L) $^3$N  $+$ \\
       &          &                 &   &  11\%  4d$^9$ 4f$^3$ ($_1^2$K) $^1$M
                                     $+$   9\%  4d$^9$ 4f$^3$ ($_1^2$L) $^3$M  \\
   177 &   9$-$   &      7.92312  & 7.96374 &    34\%  4d$^9$ 4f$^3$ ($_1^2$L) $^3$N
                                     $+$  25\%  4d$^9$ 4f$^3$ ($_1^2$L) $^3$M
                                     $+$  22\%  4d$^9$ 4f$^3$ ($_1^2$K) $^3$L  $+$ \\
       &          &                 &    &  8\%  4d$^9$ 4f$^3$ ($_1^2$L) $^1$M
                                     $+$   3\%  4d$^9$ 4f$^3$ ($_1^2$K) $^3$M  \\
   181 &  10$-$   &      7.93526  & 7.97466 &     64\%  4d$^9$ 4f$^3$ ($_1^2$L) $^3$M
                                     $+$  23\%  4d$^9$ 4f$^3$ ($_1^2$L) $^1$N
                                     $+$  12\%  4d$^9$ 4f$^3$ ($_1^2$L) $^3$N  $+$ \\
       &          &                 &    &  1\%  4d$^9$ 4f$^3$ ($_1^2$K) $^3$M  \\

\end{longtable}

\end{flushleft}

\endgroup

The first excited configuration has 4 levels and arises due to the 
$4f \rightarrow 5s$ promotion from the ground configuration. The $4f5s$ 
configuration can decay only through the electric octupole transitions in a 
single configuration approximation. The mixing of configurations opens 
additional decay channels for the $4f5s$ configuration to the ground 
configuration levels. The study of expansion coefficients for atomic state 
functions of the $4f5s$ configuration reveals that the mixing mainly with the 
$4d^{9}4f^{2}5p$, $4p^{5}4f^{2}5s$, $4f5d$, $4d^{9}4f^{3}$, and 
$4d^{9}4f^{2}5f$ configurations leads to the electric dipole transitions to the 
ground configuration. The current calculations show that the percentage 
contribution of these configurations to $4f5s$ is less than 0.1\%. The largest 
radiative transition probabilities for the $4f5s \rightarrow 4f^{2}$ transitions 
are of the order of $10^{5}$ s$^{-1}$, while the electric octupole transitions 
calculated in the single-configuration approach are of the order of $10$ 
s$^{-1}$. Thus, a small admixture of few configurations has a large effect on 
the transition probabilities.


\begingroup 

\renewcommand{\arraystretch}{1.5}
\renewcommand{\tabcolsep}{1mm}

\scriptsize

\begin{flushleft}

\LTcapwidth 15cm
\begin{longtable}{r r l r l r l r l r l r}
\caption{\label{lft}
The five greatest spontaneous radiative transition probabilities $A^r$ (in 
s$^{-1}$) from each level presented in Table \ref{energy}. Arrow marks the 
final level to which the radiative transition occurs from the level specified 
in the first column. FAC $A$-values are presented under the GRASP2K 
calculations. The sum of all radiative probabilities from the corresponding 
level is given in the last column.} \\
\hline
 No & $A$ ($s^{-1}$) & final & $A$ ($s^{-1}$) & final & $A$ ($s^{-1}$) & final & $A$ ($s^{-1}$) & final & $A$ ($s^{-1}$) & final & $\sum$ $A$ ($s^{-1}$) \\ 
 &&level&&level&&level&&level&&level& \\ \hline
\endfirsthead
\caption[]{ (continued) }  \\
\hline
 No & $A$ ($s^{-1}$) & final & $A$ ($s^{-1}$) & final & $A$ ($s^{-1}$) & final & $A$ ($s^{-1}$) & final & $A$ ($s^{-1}$) & final & $\sum$ $A$ ($s^{-1}$) \\ 
 &&level&&level&&level&&level&&level& \\ \hline
\endhead

        2 & 2.926E$-$3 & $\rightarrow$    1 &           &                     &           &      &           &      &           &       & 2.926E$-$3 \\
          & 2.899E$-$3 &                    &           &                     &           &      &           &      &
&       & 2.899E$-$3\\ 
        3 & 3.675E$+$2 & $\rightarrow$    1 &           &                     &           &      &           &      &           &       & 3.675E$+$2 \\
          & 3.504E$+$2 &                    &           &                     &           &      &           &      &
&       & 3.504E$+$2 \\

        4 & 1.592E$+$2 & $\rightarrow$    1 & 7.025E$+$0 & $\rightarrow$    3 & 2.306E$-$5 & $\rightarrow$    2 &           &      &           &       & 1.662E$+$2 \\
          & 1.535E$+$2 &                              & 6.999E$+$0 &                             & 1.978E$-$5 &                             &            &     &           &       & 1.605E$+$2 \\
        5 & 1.560E$+$2 & $\rightarrow$    2 & 1.284E$+$1 & $\rightarrow$    1 & 3.921E$-$4 & $\rightarrow$    3 & 2.186E$-$4 & $\rightarrow$    4 &           &       & 1.688E$+$2 \\
          & 1.477E$+$2 &                              & 1.235E$+$1 &                             & 3.985E$-$4 &                             &                       &                             &           &       & 1.601E$+$2 \\
        6 & 2.077E$+$2 & $\rightarrow$    3 & 1.339E$-$3 & $\rightarrow$    1 & 1.263E$-$5 & $\rightarrow$    4 &           &      &           &       & 2.077E$+$2 \\
          &  2.001E$+$2&                             & 1.176E$-$3  &                            &  1.006E$-$5 &                             &           &      &            &      & 2.001E$+$2\\
        7 & 2.799E$+$2 & $\rightarrow$    5 & 2.433E$+$2 & $\rightarrow$    4 & 5.945E$+$1 & $\rightarrow$    3 & 2.227E$+$1 & $\rightarrow$    1 & 1.496E$-$3 & $\rightarrow$    6  & 6.049E$+$2 \\
            & 2.663E$+$2&                             & 2.328E$+$2 &                             & 5.729E$+$1 &                            &  2.181E$+$1 &                             & 1.459E$-$3 &                              & 5.782E$+$2\\
        8 & 3.655E$+$2 & $\rightarrow$    2 & 1.495E$+$2 & $\rightarrow$    5 & 8.129E$-$2 & $\rightarrow$    4 & 1.973E$-$2 & $\rightarrow$    1 & 2.524E$-$8 & $\rightarrow$    7  & 5.152E$+$2 \\
           & 3.543E$+$2 &                             & 1.491E$+$2 &                             & 8.269E$-$2 &                             & 2.038E$-$2 &                             &                       &                              & 5.035E$+$2\\
       10 & 4.315E$+$1 & $\rightarrow$    2 & 2.612E$+$1 & $\rightarrow$    8 & 2.435E$+$1 & $\rightarrow$    9 & 5.204E$-$1 & $\rightarrow$    5 &           &       & 9.414E$+$1 \\
           & 4.163E$+$1 &                              & 2.653E$+$1 &                             & 2.358E$+$1 &                             &  5.308E$-$1 &                            &           &       &  9.227E$+$1\\
       11 & 3.381E$+$2 & $\rightarrow$    3 & 1.335E$+$2 & $\rightarrow$    6 & 1.455E$-$1 & $\rightarrow$    1 & 1.754E$-$2 & $\rightarrow$    4 & 1.025E$-$3 & $\rightarrow$    7  & 4.718E$+$2 \\
            & 3.209E$+$2 &                             & 1.317E$+$2 &                             & 1.412E$-$1 &                             & 1.802E$-$2 &                             &  1.131E$-$3 &                             & 4.528E$+$2\\
       12 & 3.990E$+$2 & $\rightarrow$    8 & 2.145E$+$2 & $\rightarrow$    5 & 6.166E$+$1 & $\rightarrow$   10 & 1.365E$+$1 & $\rightarrow$    2 & 1.748E$-$1 & $\rightarrow$    7  & 6.891E$+$2 \\
            & 3.814E$+$2 &                             & 2.056E$+$2 &                             & 5.720E$+$1 &                               & 1.280E$+$1 &                           &  1.793E$-$1 &                              & 6.572E$+$2\\
       13 & 3.524E$+$3 & $\rightarrow$   10 &                      &                              &                       &                               &                       &                            &                        &                             & 3.524E$+$3 \\
            & 3.416E$+$3 &                              &                      &                              &                       &                               &                       &                            &                        &                             & 3.416E$+$3 \\
       44 & 3.957E$+$1 & $\rightarrow$    6 & 3.200E$+$1 & $\rightarrow$   40 & 7.251E$+$0 & $\rightarrow$    3 & 5.193E$-$2 & $\rightarrow$   11 & 6.109E$-$6 & $\rightarrow$   31  & 7.887E$+$1 \\
          & 3.546E$+$1 &                    & 4.129E$+$1 &                    & 7.656E$+$0 &                    & 1.848E$+$2 &                    & 2.819E$-$6 &                     & 8.422E$+$1 \\
       51 & 2.828E$+$1 & $\rightarrow$   44 & 5.320E$+$0 & $\rightarrow$    6 & 1.574E$-$1 & $\rightarrow$   11 & 7.000E$-$7 & $\rightarrow$   40 & 5.836E$-$7 & $\rightarrow$   45  & 3.376E$+$1 \\
          & 3.389E$+$1 &                    & 5.536E$+$0 &                    & 2.072E$-$1 &                    & 1.378E$-$7 &                    & 4.349E$-$7 &                     & 3.963E$+$1 \\
       57 & 1.205E$+$1 & $\rightarrow$   51 & 8.853E$-$8 & $\rightarrow$   44 &           &      &           &      &           &       & 1.205E$+$1 \\
           & 1.366E$+$1  &                              & 2.266E$-$8 &                             &             &     &           &      &           &       & 1.366E$+$1 \\
       62 & 2.421E$+$3 & $\rightarrow$    6 & 1.945E$+$2 & $\rightarrow$   45 & 4.283E$+$1 & $\rightarrow$   44 & 1.810E$+$1 & $\rightarrow$   54 & 1.542E$+$1 & $\rightarrow$   40  & 2.701E$+$3 \\
            & 2.353E$+$3 &                             & 2.110E$+$2  &                              & 4.416E$+$1 &                             &  1.768E$+$1 &                              & 9.384E$+$0 &                               & 2.635E$+$3 \\
       75 & 1.041E$+$2 & $\rightarrow$   62 & 6.166E$+$1 & $\rightarrow$   51 & 2.318E$+$1 & $\rightarrow$   57 & 1.728E$+$0 & $\rightarrow$   44 & 1.457E$+$0 & $\rightarrow$    6  & 1.922E$+$2 \\
            & 1.116E$+$2  &                             & 6.077E$+$1 &                              & 2.130E$+$1  &                              & 9.618E$-$1  &                             &  1.311E$+$0 &                             &  1.959E$+$2 \\
       82 & 3.122E$+$2 & $\rightarrow$   44 & 1.291E$+$2 & $\rightarrow$   51 & 1.761E$+$1 & $\rightarrow$   11 & 2.473E$+$0 & $\rightarrow$   62 & 9.293E$-$1 & $\rightarrow$    6  & 4.629E$+$2 \\
           &  3.472E$+$2 &                               & 1.378E$+$2 &                             & 1.813E$+$1 &                              & 2.113E$+$0 &                              & 8.283E$-$1 &                              & 5.061E$+$2 \\
       85 & 2.061E$+$2 & $\rightarrow$   57 & 1.624E$+$2 & $\rightarrow$   51 & 2.721E$-$1 & $\rightarrow$   75 & 1.059E$-$1 & $\rightarrow$   82 & 3.139E$-$4 & $\rightarrow$   44  & 3.689E$+$2 \\
            & 2.171E$+$2 &                              & 1.767E$+$2 &                              & 3.034E$-$1 &                              & 1.680E$-$1 &                              & 3.198E$-$4 &                               & 3.943E$+$2 \\
       95 & 1.300E$+$3 & $\rightarrow$   11 & 1.021E$+$2 & $\rightarrow$   40 & 9.784E$+$1 & $\rightarrow$    6 & 9.239E$+$1 & $\rightarrow$   62 & 4.878E$+$1 & $\rightarrow$   45  & 1.729E$+$3 \\
            & 1.269E$+$3 &                              & 1.040E$+$2 &                              & 1.499E$+$2 &                             & 1.008E$+$2 &                              & 7.195E$+$1 &                               & 1.696E$+$3 \\
      112 & 7.582E$+$1 & $\rightarrow$   85 & 5.346E$+$1 & $\rightarrow$   57 & 4.544E$-$3 & $\rightarrow$   51 & 4.717E$-$4 & $\rightarrow$   82 & 1.028E$-$4 & $\rightarrow$   75  & 1.293E$+$2 \\
            & 8.262E$+$1  &                              & 6.198E$+$1 &                              & 5.056E$-$3 &                              & 5.441E$-$4 &                              & 1.203E$-$4 &                               &  1.446E$+$2 \\
      117 & 9.087E$+$1 & $\rightarrow$   82 & 7.448E$+$1 & $\rightarrow$   76 & 4.519E$+$1 & $\rightarrow$   75 & 4.048E$+$1 & $\rightarrow$   57 & 2.937E$+$1 & $\rightarrow$   95  & 3.237E$+$2 \\
            & 9.642E$+$1 &                               & 8.314E$+$1 &                              & 4.997E$+$1 &                              & 3.868E$+$1 &                              & 3.216E$+$1 &                               & 3.004E$+$2 \\
      121 & 2.336E$+$2 & $\rightarrow$   85 & 1.947E$+$2 & $\rightarrow$   82 & 6.721E$+$1 & $\rightarrow$   51 & 3.276E$+$1 & $\rightarrow$   57 & 4.876E$+$0 & $\rightarrow$   75  & 5.363E$+$2 \\
            & 2.551E$+$2  &                             &  2.221E$+$2 &                              & 7.342E$+$1 &                              & 3.409E$+$1 &                              & 6.187E$+$0 &                               & 5.909E$+$2 \\
      127 & 1.377E$+$3 & $\rightarrow$    6 & 4.473E$+$2 & $\rightarrow$   11 & 2.281E$+$2 & $\rightarrow$   82 & 1.129E$+$2 & $\rightarrow$   68 & 1.104E$+$2 & $\rightarrow$   54  & 2.727E$+$3 \\
            & 1.477E$+$3  &                             & 3.980E$+$2 &                              & 2.092E$+$2 &                              & 1.229E$+$2 &                              & 1.323E$+$2 &                              & 2.339E$+$3 \\
      133 & 2.872E$+$2 & $\rightarrow$    6 & 2.630E$+$2 & $\rightarrow$   76 & 6.486E$+$1 & $\rightarrow$   75 & 5.281E$+$1 & $\rightarrow$   11 & 4.568E$+$1 & $\rightarrow$   62  & 8.413E$+$2 \\
            & 1.936E$+$2  &                             & 2.723E$+$2 &                              & 6.822E$+$1 &                              & 4.897E$+$1 &                              & 5.158E$+$1 &                               & 6.347E$+$2 \\
      152 & 4.371E$+$2 & $\rightarrow$   57 & 3.857E$+$2 & $\rightarrow$   85 & 2.332E$+$2 & $\rightarrow$   75 & 1.161E$+$2 & $\rightarrow$   62 & 9.658E$+$1 & $\rightarrow$    6  & 1.430E$+$3 \\
            & 4.649E$+$2  &                              & 3.677E$+$2 &                              & 2.449E$+$2 &                              &  1.286E$+$2&                              & 9.759E$+$1 &                              & 1.304E$+$3 \\
      177 & 6.584E$+$2 & $\rightarrow$   85 & 4.430E$+$2 & $\rightarrow$  121 & 2.340E$+$2 & $\rightarrow$  127 & 1.917E$+$2 & $\rightarrow$   82 & 9.807E$+$1 & $\rightarrow$  152  & 2.045E$+$3 \\
            & 6.559E$+$2 &                              &  4.038E$+$2 &                               & 2.572E$+$2 &                               & 1.999E$+$2 &                              & 1.069E$+$2 &                                & 1.624E$+$3 \\
      181 & 5.043E$+$2 & $\rightarrow$  112 & 7.410E$+$1 & $\rightarrow$  117 & 5.224E$+$1 & $\rightarrow$  152 & 2.362E$+$1 & $\rightarrow$   51 & 2.343E$+$1 & $\rightarrow$   82  & 7.217E$+$2 \\
             & 4.797E$+$2 &                               & 8.076E$+$1 &                               & 5.652E$+$1 &                              & 2.538E$+$1 &                              & 2.233E$+$1 &                               & 6.647E$+$2 \\

\hline
\end{longtable}
\end{flushleft}

\endgroup


\begin{figure}
 \includegraphics[scale=0.50]{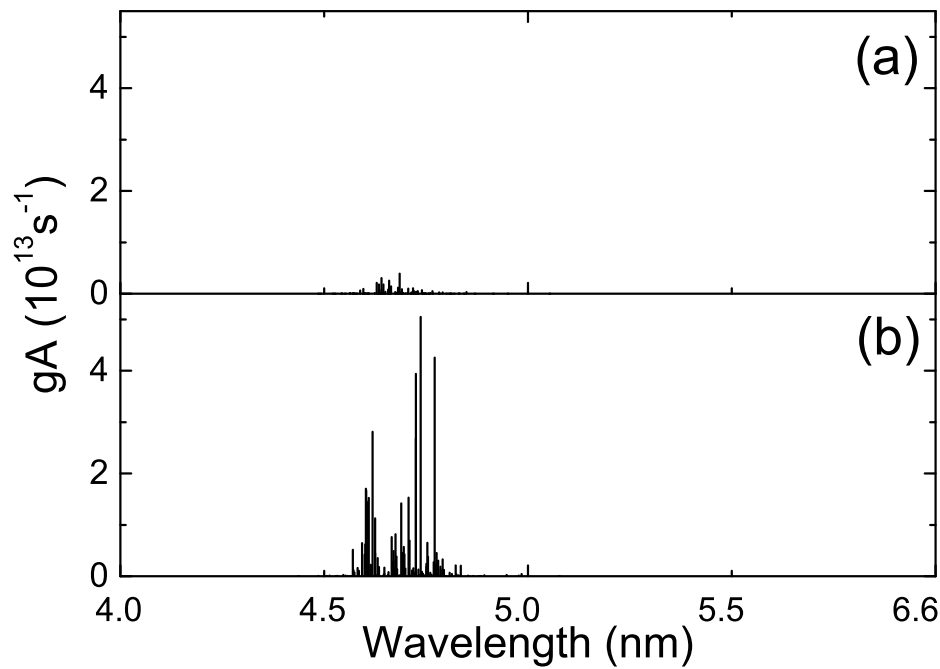}%
 \caption{
\label{ci_4f5d_4f2} 
Transition data calculated using a) single-configuration and b) configuration 
mixing methods for the $4f5d \rightarrow 4f^{2}$ transition in W$^{26+}$. 
}
\end{figure}

With an aim to estimate the importance of the correlation effects for the 
configurations forming the strongest lines in the modeled spectra, we have 
calculated the CIS values. Modeling of spectral lines is investigated in the 
next section. The CIS values for the important configurations are presented in 
Table~\ref{cis}. The current calculations for the W$^{26+}$ ion show that the 
largest CIS value corresponds to the $4f5d$ configuration interaction with the 
$4d^{9}4f^{3}$ configuration. Thus, mixing between these configurations is very 
important when the correlation effects for the  $4f5d$ configuration are 
analyzed. On the other hand, the CIS data show that the $4d^{9}4f^{3}$ 
configuration has the largest mixing with the $4d^{7}4f^{5}$ 
($T/g_{1} = 9.15\times 10^{-3}$) and $4p^{5}4d^{9}4f^{4}$ 
($T/g_{1} = 3.64\times 10^{-3}$) configurations. Influence of the $4f5d$ 
configuration is somewhat smaller compared to these two configurations 
($T/g_{1} = 2.67\times 10^{-3}$). Figure~\ref{ci_4f5d_4f2} demonstrates 
influence of the correlation effects on the $4f5d \rightarrow 4f^{2}$ 
transitions. Extended basis of configurations increases radiative transition 
probabilities for these transitions by an order of magnitude. This result 
demonstrates importance of the correlation effects for the spectrum of the 
W$^{26+}$ ion. The same effect for the $4f^{2}5d \rightarrow 4f^{3}$ transitions 
was determined for the  W$^{25+}$ ion \cite{2014jqsrt_136_108_alkauskas}.

\begingroup 

\renewcommand{\arraystretch}{1.2}
\renewcommand{\tabcolsep}{1mm}
\scriptsize

\begin{flushleft}

\renewcommand{\arraystretch}{1.5}

\LTcapwidth 15cm

\begin{longtable}{ l l l l l l l l l l l l }

\caption{\label{cis}  Configuration interaction strengths $T$  between the 
initial $K_{1}$ and admixed $K_{2}$ configurations divided by the statistical 
weight $g_{1}$ of the initial configuration for some configurations of the 
$W^{26+}$ ion. Occupation numbers of shells of admixed configurations  $K_{2}$ 
are given relatively to the corresponding initial configuration $K_{1}$
(for example, in the case of $K_{1} = 4f^{2}$, notation $K_{2} = 4d^{-2}4f^{2}$ 
corresponds to $4d^{8}4f^{4}$ configuration, and in the case of 
$K_{1} = 4d^{9}4f^{3}$, $K_{2} = 4d^{-2}4f^{2}$ corresponds to $4d^{7}4f^{5}$ 
configuration). 
} \\
\hline
 $K_{1}$ & $T/g_{1}$ & $K_{2}$ && $T/g_{1}$ & $K_{2}$ && $T/g_{1}$ & $K_{2}$ && $T/g_{1}$ & $K_{2}$ \\
\hline
\endfirsthead
\caption[]{ (continued) }  \\
\hline
 $K_{1}$ & $T/g_{1}$ & $K_{2}$ && $T/g_{1}$ & $K_{2}$ && $T/g_{1}$ & $K_{2}$ && $T/g_{1}$ & $K_{2}$ \\
\hline
\endhead
\hline \multicolumn{4}{r}{\textit{Continued on next page}} \\
\endfoot
\hline
\endlastfoot
$4f^{2}$ & $1.29 {-2}$ & $4d^{-2}4f^{2}$ && $1.10 {-3}$ & $4d^{-1}5g^{1}$ && $7.39 {-4}$ & $4p^{-1}4f^{1}$ && $5.11 {-4}$ & $4p^{-2}4f^{2}$\\

& $4.57 {-4}$ & $4d^{-1} 6g^{1}$ && $3.88 {-4}$ & $3d^{-1}4d^{-1}4f^{2}$ && $3.61 {-4}$ & $4p^{-1}4d^{-1}4f^{1}5g^{1}$ && $2.64 {-4}$ & $4d^{-2}5d^{2}$\\

& $2.58 {-4}$ & $4p^{-1}4d^{-1}5p^{1}5d^{1}$ && $2.23 {-4}$ & $4d^{-1}7g^{1}$ && $1.83 {-4}$ & $4p^{-1}4d^{-1}4f^{1}6g^{1}$ && $1.78 {-4}$ & $4d^{-1}4f^{-1}5d^{1}5f^{1}$\\

& $1.69 {-4}$ & $4s^{-1}4d^{-1}4f^{2}$ && $1.30 {-4}$ & $4d^{-2}5d^{1}6d^{1}$ && $1.19 {-4}$ & $4p^{-1}4d^{-1}4f^{1}5d^{1}$ && $9.78 {-5}$ & $4p^{-1}4d^{-1}4f^{1}7g^{1}$\\

& $9.58 {-5}$ & $3d^{-2}4f^{2}$ && $8.85 {-5}$ & $3d^{-1}4d^{-1}4f^{1}5f^{1}$ && $7.89 {-5}$ & $4d^{-1}5d^{1}$ && $7.72 {-5}$ & $4p^{-1}4f^{-1}5p^{1}5f^{1}$\\

& $6.94 {-5}$ & $4s^{-1}4d^{-1}5s^{1}5d^{1}$ && $6.78 {-5}$ & $4p^{-1}4d^{-1}5p^{1}6d^{1}$ && $5.82 {-5}$ & $3d^{-2}4f^{1}5f^{1}$ &&  $5.79 {-5}$ & $4p^{-1}4d^{-1}5d^{1}6p^{1}$\\

& $5.23 {-5}$ & $4d^{-2}5d^{1}7d^{1}$ && $5.23 {-5}$ & $4s^{-1}4d^{-1}4f^{1}5p^{1}$ && $5.06 {-5}$ & $4p^{-2}5p^{2}$ && $4.45 {-5}$ & $4d^{-2}4f^{1}5p^{1}$\\

& $4.45 {-5}$ & $4d^{-1}4f^{-1}5d^{1}6f^{1}$ && $4.16 {-5}$ & $4d^{-1}4f^{-1}5f^{1}6d^{1}$\\

$4d^{9}4f^{3}$     & $9.15 {-3}$ & $4d^{-2}4f^{2}$ && $3.64 {-3}$ & $4p^{-1}4f^{1}$ && $2.67 {-3}$ & $4d^{1}4f^{-2}5d^{1}$ && $1.47 {-3}$ & $4d^{-1}5g^{1}$\\

& $6.11 {-4}$ & $4d^{-1}6g^{1}$ && $4.29 {-4}$ & $4p^{-2} 4f^{2}$ && $3.00 {-4}$ & $4p^{-1}4d^{-1}4f^{1}5g^{1}$ && $2.98 {-4}$ & $4d^{-1}7g^{1}$\\

& $2.93 {-4}$ & $3d^{-1}4d^{-1}4f^{2}$ && $2.84 {-4}$ & $4d^{1}4f^{-2}5g^{1}$ && $2.42 {-4}$ & $4d^{-1}4f^{-1}5d^{1}5f^{1}$ && $2.35 {-4}$ & $4p^{-1}4d^{-1}5p^{1}5d^{1}$\\

& $2.15 {-4}$ & $4d^{-2}5d^{2}$ && $1.52 {-4}$ & $4p^{-1}4d^{-1}4f^{1}6g^{1}$ && $1.42 {-4}$ & $4p^{-1}4d^{1}4f^{-1}5g^{1}$ && $1.26 {-4}$ & $4s^{-1}4d^{-1}4f^{2}$\\

& $1.16 {-4}$ & $4d^{-1}5d^{1}$ && $1.16 {-4}$ & $4p^{-1}4f^{-1}5p^{1}5f^{1}$ && $1.13 {-4}$ & $4s^{-1}4d^{1}$ && $1.05 {-4}$ & $4d^{-2}5d^{1}6d^{1}$\\

& $1.01 {-4}$ & $4p^{-1}4d^{-1}4f^{1}5d^{1}$ && $8.08 {-5}$ & $4p^{-1}4d^{-1}4f^{1}7g^{1}$ && $7.96 {-5}$ & $4s^{-1}4p^{-1}4d^{1}4f^{1}$ && $7.94 {-5}$ & $3d^{-2}4f^{2}$\\

& $7.32 {-5}$ & $3d^{-1}4d^{-1}4f^{1}5f^{1}$ && $6.32 {-5}$ & $4s^{-1}4d^{-1}5s^{1}5d^{1}$ && $6.16 {-5}$ & $4p^{-1}4d^{-1}5p^{1}6d^{1}$ && $6.04 {-5}$ & $4d^{-1}4f^{-1}5d^{1}6f^{1}$\\

& $5.75 {-5}$ & $4p^{-1}4d^{1}4f^{-1}6g^{1}$ && $5.62 {-5}$ & $4d^{-1}4f^{-1}5f^{1}6d^{1}$\\

$4f 5s$ & $1.51 {-2}$ & $4d^{-2}4f^{2}$ && $5.23 {-3}$ & $4d^{-1}4f^{1}5s^{-1}5p^{1}$ && $6.06 {-4}$ & $4p^{-2}4f^{2}$ && $5.94 {-4}$ & $4d^{-1}5g^{1}$\\

& $4.67 {-4}$ & $3d^{-1}4d^{-1}4f^{2}$ && $4.43 {-4}$ & $4p^{-1}4f^{1}$ && $4.02 {-4}$ & $4p^{-1}4d^{-1}4f^{1}5g^{1}$ && $2.88 {-4}$ & $5s^{-1}5d^{1}$\\

& $2.60 {-4}$ & $4d^{-2}5d^{2}$ && $2.57 {-4}$ & $4p^{-1}4d^{-1}5p^{1}5d^{1}$ && $2.42 {-4}$ & $4d^{-1}6g^{1}$ && $2.13 {-4}$ & $4d^{-1}4f^{1}5s^{-1}6p^{1}$\\

& $2.01 {-4}$ & $4s^{-1}4d^{-1}4f^{2}$ && $1.98 {-4}$ & $4p^{-1}4d^{-1}4f^{1}6g^{1}$ && $1.80 {-4}$ & $4p^{-1}4f^{1}5s^{-1}5d^{1}$ && $1.27 {-4}$ & $4d^{-2}5d^{1}6d^{1}$\\

& $1.23 {-4}$ & $4p^{-1}4d^{-1}4f^{1}5d^{1}$ && $1.18 {-4}$ & $4d^{-1}7g^{1}$ && $1.17 {-4}$ & $3d^{-2}4f^{2}$ && $1.05 {-4}$ & $4f^{-1}5s^{-1}5p^{1} 5g^{1}$\\

& $1.05 {-4}$ & $4p^{-1}4d^{-1}4f^{1}7g^{1}$ && $9.55 {-5}$ & $3d^{-1}4d^{-1}4f^{1}5f^{1}$ && $9.46 {-5}$ & $4d^{-1}5s^{-1}5d^{1}5g^{1}$ && $8.74 {-5}$ & $4d^{-1}4f^{-1}5d^{1}5f^{1}$\\

& $6.80 {-5}$ & $4d^{-1}4f^{1}5s^{-1}5f^{1}$ && $6.65 {-5}$ & $4p^{-1}4d^{-1}5p^{1}6d^{1}$ && $6.49 {-5}$ & $3d^{-2}4f^{1}5f^{1}$ && $6.36 {-5}$ & $4d^{-1}5d^{1}$\\

& $5.68 {-5}$ & $4s^{-1}4d^{-1}4f^{1}5p^{1}$ && $5.66 {-5}$ & $4p^{-1}4d^{-1}5d^{1}6p^{1}$\\

$4f 5p$ & $1.51 {-2}$ & $4d^{-2}4f^{2}$ && $4.51 {-3}$ & $4d^{-1}4f^{1}5p^{-1}5s^{1}$ && $2.74 {-3}$ & $4d^{-1}4f^{1}5p^{-1}5d^{1}$ && $6.52 {-4}$ & $4d^{-1}5g^{1}$\\

& $6.07 {-4}$ & $4p^{-2}4f^{2}$ && $5.75 {-4}$ & $4p^{-1}4f^{1}$ && $4.67 {-4}$ & $3d^{-1}4d^{-1}4f^{2}$ && $4.04 {-4}$ & $4p^{-1}4d^{-1}4f^{1}5g^{1}$\\

& $2.60 {-4}$ & $4d^{-2}5d^{2}$ && $2.60 {-4}$ & $4d^{-1}6g^{1}$ && $2.13 {-4}$ & $4p^{-1}4d^{-1}5p^{1}5d^{1}$ && $2.02 {-4}$ & $4s^{-1}4d^{-1}4f^{2}$\\

& $1.98 {-4}$ & $4p^{-1}4d^{-1}4f^{1}6g^{1}$ && $1.58 {-4}$ & $4d^{-1}4f^{1}5p^{-1}6d^{1}$ && $1.26 {-4}$ & $4d^{-2}5d^{1}6d^{1}$ && $1.25 {-4}$ & $4d^{-1}7g^{1}$\\

& $1.24 {-4}$ & $4p^{-1}4d^{-1}4f^{1}5d^{1}$ && $1.17 {-4}$ & $3d^{-2}4f^{2}$ && $1.05 {-4}$ & $4p^{-1}4d^{-1}4f^{1}7g^{1}$ && $1.03 {-4}$ & $5p^{-1}5f^{1}$\\

& $9.56 {-5}$ & $3d^{-1}4d^{-1}4f^{1}5f^{1}$ && $9.23 {-5}$ & $4p^{-1}4f^{1}5p^{-1}5f^{1}$ && $8.74 {-5}$ & $4d^{-1}4f^{-1}5d^{1}5f^{1}$ && $7.11 {-5}$ & $4d^{-1}5d^{1}$\\

& $6.89 {-5}$ & $4s^{-1}4d^{-1}5s^{1}5d^{1}$ && $6.50 {-5}$ & $3d^{-2}4f^{1}5f^{1}$ && $5.96 {-5}$ & $4f^{-1}5p^{-1}5d^{1}5g^{1}$ && $5.94 {-5}$ & $4f^{-1}5p^{-1}5s^{1}5g^{1}$\\

& $5.65 {-5}$ & $4p^{-1}4d^{-1}5d^{1}6p^{1}$ && $5.51 {-5}$ & $4p^{-1}4d^{-1}5p^{1}6d^{1}$\\

$4f 5d$ & $6.92 {-2}$ & $4d^{-1}4f^{2}5d^{-1}$ && $1.51 {-2}$ & $4d^{-2}4f^{2}$ && $5.88 {-3}$ & $4d^{-1}4f^{1}5p^{1}5d^{-1}$ && $2.00 {-3}$ & $4d^{-1}4f^{1}5d^{-1}5f^{1}$\\

& $6.38 {-4}$ & $4d^{-1}5g^{1}$ && $6.08 {-4}$ & $4p^{-2}4f^{2}$ && $5.42 {-4}$ & $4p^{-1}4f^{1}$ && $4.68 {-4}$ & $3d^{-1}4d^{-1}4f^{2}$\\

& $4.06 {-4}$ & $4p^{-1}4d^{-1}4f^{1}5g^{1}$ && $2.53 {-4}$ & $4d^{-1}6g^{1}$ && $2.30 {-4}$ & $4p^{-1}4d^{-1}5p^{1}5d^{1}$ && $2.09 {-4}$ & $4d^{-2}5d^{2}$\\

& $2.02 {-4}$ & $4s^{-1}4d^{-1}4f^{2}$ && $1.98 {-4}$ & $4p^{-1}4d^{-1}4f^{1}6g^{1}$ && $1.30 {-4}$ & $4d^{-1}4f^{1}5d^{-1}6f^{1}$ && $1.30 {-4}$ & $4p^{-1}4f^{1}5d^{-1}5s^{1}$\\

& $1.22 {-4}$ & $4d^{-1}7g^{1}$ && $1.17 {-4}$ & $3d^{-2}4f^{2}$ && $1.14 {-4}$ & $4d^{-2}5d^{1}6d^{1}$ && $1.10 {-4}$ & $4p^{-1}4d^{-1}4f^{1}5d^{1}$\\

& $1.05 {-4}$ & $4p^{-1}4d^{-1}4f^{1}7g^{1}$ && $9.59 {-5}$ & $3d^{-1}4d^{-1}4f^{1}5f^{1}$ && $7.88 {-5}$ & $4d^{-1}4f^{-1}5d^{1}5f^{1}$ && $7.07 {-5}$ & $4f^{-1}5g^{1}$\\

& $6.58 {-5}$ & $4p^{-1}4d^{-1}5p^{1}6d^{1}$ && $6.52 {-5}$ & $3d^{-2}4f^{1}5f^{1}$ && $6.19 {-5}$ & $4s^{-1}4d^{-1}5s^{1}5d^{1}$ && $5.78 {-5}$ & $4d^{-1}4f^{1}5d^{-1}6p^{1}$\\

& $5.76 {-5}$ & $5d^{-1}5s^{1}$ && $5.54 {-5}$ & $4s^{-1}4d^{-1}4f^{1}5p^{1}$\\

$4f 5f$ & $2.35 {-1}$ & $5s^{-1}5f^{-1}5p^{1}5d^{1}$ && $1.50 {-2}$ & $4d^{-2}4f^{2}$ && $9.84 {-3}$ & $4d^{-1}4f^{1}5s^{-1}5p^{1}$ && $8.48 {-3}$ & $5s^{-2}5p^{2}$\\

& $6.96 {-3}$ & $4d^{-1}4f^{1}5f^{-1}5d^{1}$ && $4.18 {-3}$ & $5s^{-1}5f^{-1}5p^{1}5g^{1}$ && $3.71 {-3}$ & $4f^{-1}5f^{-1}5p^{2}$ && $1.24 {-3}$ & $5s^{-1} 5d^{1}$\\

& $9.60 {-4}$ & $4f^{1}5s^{-1}5f^{-1}7s^{1}$ && $8.46 {-4}$ & $4d^{-1}4f^{1}5f^{-1}5g^{1}$ && $6.03 {-4}$ & $4p^{-2}4f^{2}$ && $5.94 {-4}$ & $4d^{-1}5g^{1}$\\

& $5.32 {-4}$ & $4p^{-1}4f^{1}5f^{-1}5p^{1}$ && $5.01 {-4}$ & $4p^{-1}4f^{1}$ && $4.89 {-4}$ & $5s^{-2}5d^{2}$ && $4.74 {-4}$ & $4f^{1}5s^{-1}5f^{-1}6s^{1}$\\

& $4.60 {-4}$ & $3d^{-1}4d^{-1}4f^{2}$ && $4.31 {-4}$ & $4d^{-1}4f^{1}5s^{-1}6p^{1}$ && $3.60 {-4}$ & $4p^{-1}4d^{-1}4f^{1}5g^{1}$ && $3.44 {-4}$ & $4p^{-1}4f^{1} 5s^{-1}5d^{1}$\\

& $3.34 {-4}$ & $4f^{1}5s^{-1}5f^{-1}5g^{1}$ && $2.58 {-4}$ & $4d^{-2}5d^{2}$ && $2.55 {-4}$ & $4p^{-1}4d^{-1}5p^{1}5d^{1}$ && $2.44 {-4}$ & $4d^{-1}4f^{1}5f^{-1}6d^{1} $\\

& $2.40 {-4}$ & $4d^{-1}6g^{1}$ && $1.99 {-4}$ & $4s^{-1}4d^{-1}4f^{2}$ && $1.95 {-4}$ & $4f^{-1}5s^{-1}5p^{1}5g^{1}$ && $1.83 {-4}$ & $4p^{-1}4d^{-1}4f^{1}6g^{1}$\\

& $1.83 {-4}$ & $4f^{1}5s^{-2}5f^{1}$ && $1.71 {-4}$ & $4d^{-1}5s^{-1}5d^{1}5g^{1}$\\

$4f 5g$ & $1.52 {-2}$ & $4d^{-2}4f^{2}$ && $4.11 {-3}$ & $4d^{-1}4f^{2}5g^{-1}$ && $4.00 {-3}$ & $4d^{-1}4f^{1}5g^{-1}5p^{1}$ && $1.96 {-3}$ & $4d^{-1}4f^{1}5g^{-1}5f^{1}$\\

& $6.08 {-4}$ & $4p^{-2}4f^{2}$ && $6.02 {-4}$ & $4d^{-1}5g^{1}$ && $5.35 {-4}$ & $4p^{-1}4f^{1}$ && $4.68 {-4}$ & $3d^{-1}4d^{-1}4f^{2}$\\

& $4.67 {-4}$ & $4d^{-1}4f^{1}5g^{-1}6h^{1}$ && $2.58 {-4}$ & $4d^{-2}5d^{2}$ && $2.56 {-4}$ & $4d^{-1}6g^{1}$ && $2.53 {-4}$ & $4p^{-1}4d^{-1}5p^{1}5d^{1}$\\

& $2.02 {-4}$ & $4s^{-1}4d^{-1}4f^{2}$ && $1.94 {-4}$ & $5g^{-1}6g^{1}$ && $1.79 {-4}$ & $4d^{-1}4f^{1}5g^{-1}7h^{1}$ && $1.26 {-4}$ & $4d^{-2}5d^{1}6d^{1}$\\

& $1.23 {-4}$ & $4f^{-1}5g^{-1}5p^{1}5d^{1}$ && $1.22 {-4}$ & $4d^{-1}7g^{1}$ && $1.22 {-4}$ & $4p^{-1}4d^{-1}4f^{1}5d^{1}$ && $1.18 {-4}$ & $3d^{-2}4f^{2}$\\

& $9.56 {-5}$ & $3d^{-1}4d^{-1}4f^{1}5f^{1}$ && $8.67 {-5}$ & $4d^{-1}4f^{-1}5d^{1}5f^{1}$ && $8.27 {-5}$ & $4p^{-1}4f^{1}5g^{-1}5d^{1}$ && $8.04 {-5}$ & $4d^{-1}5d^{1}$\\

& $7.41 {-5}$ & $4d^{-1}5g^{-1}5p^{1}5f^{1}$ && $6.83 {-5}$ & $4s^{-1}4d^{-1}5s^{1}5d^{1}$ && $6.58 {-5}$ & $4p^{-1}4d^{-1}5p^{1}6d^{1}$ && $6.53 {-5}$ & $3d^{-2}4f^{1}5f^{1}$\\

& $6.26 {-5}$ & $4s^{-1}4f^{2}5g^{-1}$ && $5.65 {-5}$ & $4p^{-1}4d^{-1}5d^{1}6p^{1}$\\

$4f 6g$ & $1.52 {-2}$ & $4d^{-2}4f^{2}$ && $8.00 {-4}$ & $4d^{-1}4f^{2}6g^{-1}$ && $7.12 {-4}$ & $4p^{-1}4f^{1}$ && $6.42 {-4}$ & $4s^{-1}4f^{2}6g^{-1}$\\

& $6.24 {-4}$ & $4d^{-1}5g^{1}$ && $6.09 {-4}$ & $4p^{-2}4f^{2}$ && $4.74 {-4}$ & $4d^{-1}4f^{1}6g^{-1}6f^{1}$ && $4.69 {-4}$ & $4p^{-1}4f^{1}$\\

& $4.69 {-4}$ & $3d^{-1}4d^{-1}4f^{2}$ && $2.60 {-4}$ & $4d^{-2}5d^{2}$ && $2.55 {-4}$ & $4p^{-1}4d^{-1}5p^{1}5d^{1}$ && $2.32 {-4}$ & $4d^{-1}6g^{1}$\\

& $2.03 {-4}$ & $4s^{-1}4d^{-1}4f^{2}$ && $1.94 {-4}$ & $5g^{1}6g^{-1}$ && $1.71 {-4}$ & $6g^{-1}7g^{1}$ && $1.27 {-4}$ & $4d^{-2}5d^{1}6d^{1}$\\

& $1.21 {-4}$ & $4p^{-1}4d^{-1}4f^{1}5d^{1}$ && $1.18 {-4}$ & $3d^{-2}4f^{2}$ && $1.17 {-4}$ & $4d^{-1}7g^{1}$ && $9.67 {-5}$ & $3d^{-1}4d^{-1}4f^{1}5f^{1}$\\

& $8.73 {-5}$ & $4d^{-1}5d^{1}4f^{-1}5f^{1}$ && $8.31 {-5}$ & $4d^{-1}4f^{1}6g^{-1}7h^{1}$ && $6.89 {-5}$ & $4s^{-1}4d^{-1}5s^{1}5d^{1}$ && $6.64 {-5}$ & $3d^{-2}4f^{1}5f^{1}$\\

& $6.62 {-5}$ & $4p^{-1}4d^{-1}5p^{1}6d^{1}$ && $5.70 {-5}$ & $4p^{-1}4d^{-1}5d^{1}6p^{1}$ && $5.51 {-5}$ & $4s^{-1}4d^{-1}4f^{1}5p^{1}$ && $5.06 {-5}$ & $4d^{-1}4f^{1}5f^{1}6g^{-1}$\\

& $5.05 {-5}$ & $4d^{-2}5d^{1}7d^{-1}$ && $5.02 {-5}$ & $4p^{-2}5p^{2}$\\

$4f 7g$ & $1.52 {-2}$ & $4d^{-2}4f^{2}$ && $9.07 {-3}$ & $4d^{-1}4f^{1}5f^{1}7g^{-1}$ && $6.25 {-4}$ & $4s^{-1}4f^{2}7g^{-1}$ && $6.18 {-4}$ & $4d^{-1}5g^{1}$\\

& $6.09 {-4}$ & $4p^{-2}4f^{2}$ && $5.79 {-4}$ & $4d^{-1}5s^{1}5d^{1}7g^{-1}$ && $4.69 {-4}$ & $3d^{-1}4d^{-1}4f^{2}$ && $4.54 {-4}$ & $4p^{-1}4f^{1}$\\

& $2.96 {-4}$ & $4d^{-1}4f^{2}7g^{-1}$ && $2.60 {-4}$ & $4d^{-2}5d^{2}$ && $2.55 {-4}$ & $4p^{-1}4d^{-1}5p^{1}5d^{1}$ && $2.47 {-4}$ & $4d^{-1}6g^{1}$\\

& $2.03 {-4}$ & $4s^{-1}4d^{-1}4f^{2}$ && $1.71 {-4}$ & $6g^{1}7g^{-1}$ && $1.62 {-4}$ & $4d^{-1}4f^{1}7f^{1}7g^{-1}$ && $1.27 {-4}$ & $4d^{-2}5d^{1}6d^{1}$\\

& $1.25 {-4}$ & $4d^{-1}4f^{1}6p^{1}7g^{-1}$ && $1.20 {-4}$ & $4p^{-1}4d^{-1}4f^{1}5d^{1}$ && $1.19 {-4}$ & $4p^{-1}4f^{1}5d^{1}7g^{-1}$ && $1.18 {-4}$ & $3d^{-2}4f^{2}$\\

& $1.10 {-4}$ & $4d^{-1}7g^{1}$ && $9.70 {-5}$ & $3d^{-1}4d^{-1}4f^{1}5f^{1}$ && $9.53 {-5}$ & $4d^{-1}4f^{1}6f^{1}7g^{-1}$ && $8.97 {-5}$ & $4f^{-1}5p^{1}5g^{1}7g^{-1}$\\

& $8.74 {-5}$ & $4d^{-1}5d^{1}4f^{-1}5f^{1}$ && $6.90 {-5}$ & $4s^{-1}4d^{-1}5s^{1}5d^{1}$ && $6.67 {-5}$ & $3d^{-2}4f^{1}5f^{1}$ && $6.65 {-5}$ & $4p^{-1}4d^{-1}5p^{1}6d^{1}$\\

& $5.72 {-5}$ & $4p^{-1}4d^{-1}5d^{1}6p^{1}$ && $5.51 {-5}$ & $4s^{-1}4d^{-1}4f^{1}5p^{1}$\\
\end{longtable}

\end{flushleft}

\endgroup


It has to be noted that for all presented configurations, the core correlations 
corresponding to promotions from the $3s$ and $3d$ shells have to be taken into 
account even for the highly excited configurations. Here we consider 
$1s^{2}2s^{2}2p^{6}3s^{2}3p^{6}3d^{10}4s^{2}4p^{6}4d^{10}$ as the core shells 
and $4f$, $5l$ ($l=0, 1, \ldots, 4$), $6g$, and $7g$ as the valence shells. 
However, the $T/g_{1}$ values for these correlations are approximately by two 
orders of magnitude smaller compared with the configurations having the largest 
impact. In the discussion, we use the terms 'core', 'core-core', 'valence', and 
'valence-valence' correlations, meaning correlations with configurations 
involving promotion of one or two electrons from the core or valence shells, 
respectively. For all presented configurations, the $4d^{-2}4f^{2}$ correlation 
is the strongest one except for the $4f5d$ case where the $4d^{-1}4f^{2}5d^{-1}$ 
correlation dominates. The core and core-core correlations play the major role 
compared to the valence and valence-valence correlations.  

The magnetic dipole and electric quadrupole transitions among the levels of the 
ground configuration of the W$^{26+}$ ion have been studied using the MCDF 
approach \cite{2011jpb_44_145004_ding}. Recently, the investigation has been 
extended by the MR-RMBT calculations using FAC and by the MCDF calculations 
using GRASP2K for all levels of the ground configuration 
\cite{2014pra_90_052517_fei}. Good agreement with experiment is {\bf achieved} in both 
works. On the other hand, these investigations have used the MR-RMBPT 
\cite{2014pra_90_052517_fei} and the extended optimal level (EOL) approximation 
in the MCDF calculations \cite{2011jpb_44_145004_ding, 2014pra_90_052517_fei}, 
which provide accurate data only for few defined levels. We investigate a much 
larger group of levels and transitions among them, therefore, the extended 
average level (EAL) approximation is employed in our MCDF studies. For example, 
the final wavelength for the ${}^{3}H_{5} \rightarrow {}^{3}H_{4}$ transition 
equals to $388.43$ nm in \cite{2011jpb_44_145004_ding} and $390.9$ nm in 
\cite{2014pra_90_052517_fei}, while our MCDF calculation gives $399.05$ nm, and 
our FAC calculation gives $405.55$ nm. Our numbers are significantly larger 
compared with the experimental value of $389.41$ nm. This suggests that a much 
larger basis is needed to achieve good agreement with experiment for transitions 
among the levels of the ground configuration.

\section{The modeling of emission spectra}

Three regions of wavelengths can be highlighted in the modeled spectra of the 
W$^{26+}$ ion. The first region from around 1.5 nm to 4 nm corresponds mainly 
to the transitions from $4fng$ to the ground configuration. Transitions from 
the $4d^{9} 4f^{3}$ and $4f5d$ configurations concentrate in the $4-7$ nm region 
where the largest peak of emission from various tungsten spectra is located. 
Lines in the third region ($10-30$ nm) originate from the ($n=5$)$-$($n=5$) and 
$4f5s \rightarrow 4f^{2}$ transitions. Our study of spectra for the 
W$^{26+}$ ion is presented in these wavelength intervals. Unfortunately, there 
are no published results of EBIT measurements in the third region.

\subsection{$1.5-4$ nm region}


\begin{figure}
 \includegraphics[scale=0.35]{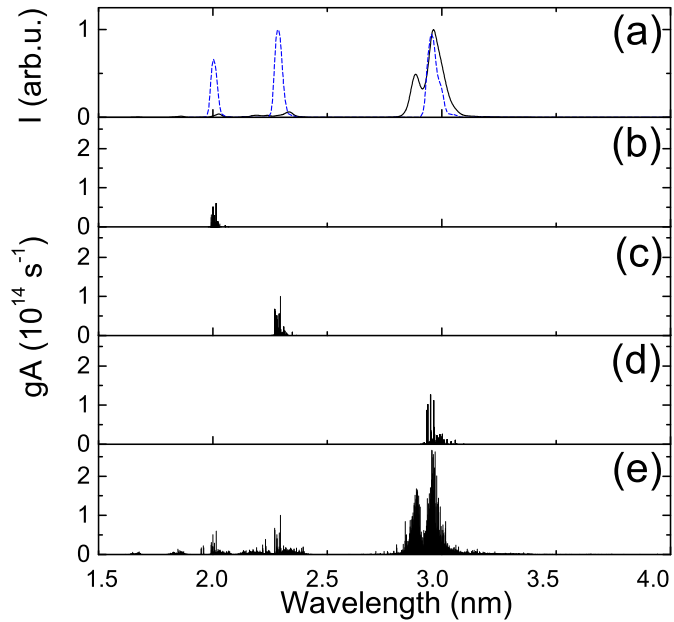}%
 \caption{
\label{gA_1_4} 
Calculated transition data for W$^{26+}$ in the $1.5- 4$ nm range;
(a) convoluted $gA$ spectra with a full width at half maximum  of 0.02 nm. 
The dashed line (blue online) corresponds to the sum of contributions from 
(b) $4f7g \rightarrow 4f^{2}$, 
(c) $4f6g \rightarrow 4f^{2}$, and 
(d) $4f5g \rightarrow 4f^{2}$ transitions. 
The solid line corresponds to 
(e) the spectrum of all transitions considered in this work.
}
\end{figure}


The wavelength region of $1.5-4$ nm is covered by lines from the 
$4f5g \rightarrow 4f^{2}$, $4f6g \rightarrow 4f^{2}$, $4f7g \rightarrow 4f^{2}$, 
$4d^{9}4f^{2}5p \rightarrow 4f^{2}$, $4p^{5}4f^{2}5d \rightarrow 4f^{2}$,  
$4p^{5}4f^{2}5s \rightarrow 4f^{2}$, and $4d^{9}4f^{2}5f \rightarrow 4f^{2}$ 
transitions. Transitions from the $4fng$ ($n=5, 6, 7$) configurations are 
presented in Fig.~\ref{gA_1_4}. It was determined for the W$^{25+}$ ion that 
the $4f^{2}5g \rightarrow 4f^{3}$ and $4f^{2}6g \rightarrow 4f^{3}$ transitions 
are the most important in this region \cite{2014jqsrt_136_108_alkauskas}.


\begin{figure}
 \includegraphics[scale=0.45]{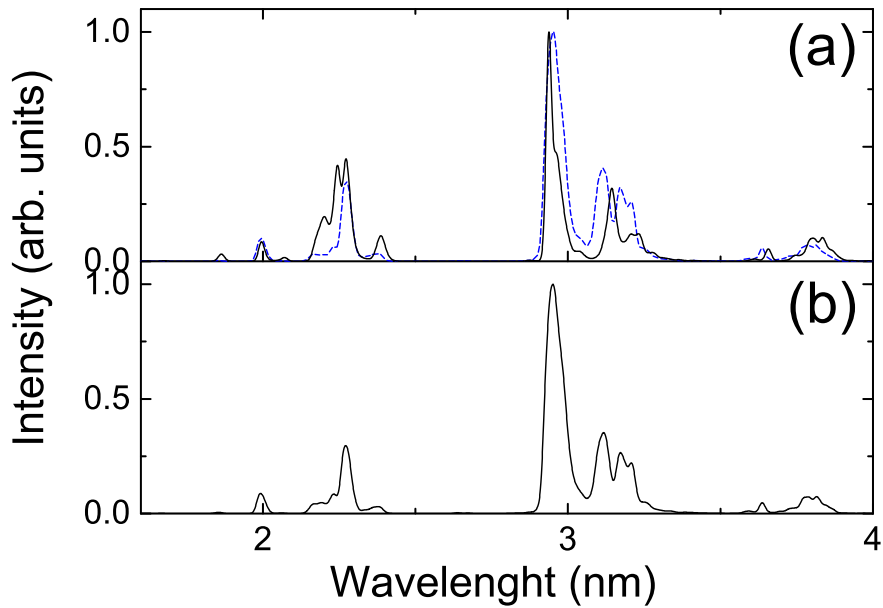}%
 \caption{
\label{w26_corona_1_4} 
Spectra corresponding to (a) corona models and (b) CRM in the $1.5-4$ nm range. 
For the corona models, the solid line represents spectrum simulated in an 
approach where the electric dipole line strengths are used instead of the 
electron-impact excitation rates; the dashed line (blue online) - the 
electron-impact excitation rates calculated using the DW method.
}
\end{figure}


As mentioned above, two approaches for the electron-impact excitation rates are 
used in the corona modeling. One can see from Fig.~\ref{w26_corona_1_4} that 
a very similar group of lines appears in both models. Good agreement with the 
CRM spectrum means that modeling the line intensities only with the electric 
dipole line strengths instead of the electron-impact excitation rates is 
justified to determine the strongest lines in a low density plasma of the 
W$^{26+}$ ion.  

Podpaly et al. \cite{2011cjp_89_591_podpaly} and Chowdhuri et al. 
\cite{2007pfr_2_s1060_chowdhuri} observed fusion spectra of tungsten in this 
region. Our data shows that some unidentified lines in their measurements can 
belong to the W$^{26+}$ ion. The strongest group of lines in the region 
originates from the $4f5g \rightarrow 4f^{2}$ transitions with a peak at 
$2.943$ nm. Podpaly et al. \cite{2011cjp_89_591_podpaly} observed a line with 
the wavelength of $2.977\pm0.002$ nm in a fusion plasma but the line 
identification was not proposed. The same line was measured at $2.951\pm 0.003$ 
nm by Chowdhuri et al. \cite{2007pfr_2_s1060_chowdhuri}. The identification of 
this line was suggested by Sugar et al. \cite{1980pra_21_2096_sugar} as the 
$4d^{10}\: {}^{1}S_{0} - 4d^{9} 5p\: {}^{1}P_{1}$ transition in W$^{28+}$. 
This wavelength was determined from the extrapolation of Pd isoelectronic 
sequence resonance lines \cite{1980pra_21_2096_sugar}. Lines in the region 
observed in the CoBIT and LHD spectra were attributed to the $4f5g - 4f^{2}$ 
transition \cite{2012aipcp_1438_91_sakaue, 2013aipcp_1545_143_morita}. 
It was previously also proposed that this line could belong to the
$4f5g - 4f^{2}$ transition in the W$^{26+}$ ion \cite{2010jpb_43_0953_harte, 
2012jpb_45_205002_harte}. A pseudorelativistic approach with scaled integrals 
\cite{Cowan81} has been used in the latter studies. Their intensity-weighted 
mean wavelength of $2.97$ nm for the $gA$ (where $g$ is statistical weight of 
the initial level, and $A$ is transition probability) spectrum is in good 
agreement with the experimental value \cite{2010jpb_43_0953_harte}. However, 
the distribution of the $gA$ values does not fully determine the spectral shape 
and dominant transitions since the population mechanisms are not taken into 
account. In addition, the mean wavelength value is shifted compared to the peak 
value due to asymmetric distribution of lines. For example, the peak of the 
$4f5g \rightarrow 4f^{2}$ transition in the $gA$ spectrum appears at $2.956$ 
nm in our calculations while the $gA$-weighted mean wavelength corresponds to 
$2.966$ nm. Unfortunately, the current modeling does not allow us to assess
which of the ions are responsible for the emission in this region. On the other 
hand, the wide peak in the experimental spectrum \cite{2011cjp_89_591_podpaly} 
suggests that the emission from more than one ion is observed. 

An additional peak at $3.141$ nm formed by the 
$4d^{9}4f^{2}5p \rightarrow 4f^{2}$ transitions is seen in the  spectrum modeled 
with the MCDF data. A similar value of $3.120$ nm is obtained from the FAC 
calculations. The pseudorelativistic approach yielded $3.12$ nm weighted mean 
wavelength of the $gA$ spectrum \cite{2010jpb_43_0953_harte}. Podpaly et al.  
\cite{2011cjp_89_591_podpaly} observed lines in the same region with a peak at 
$3.145\pm 0.003$ nm. However, intensities of these lines in our calculations are 
few times weaker than the intensities of the $4f5g \rightarrow 4f^{2}$ 
transitions in the $3$ nm region. Therefore, the contribution of the 
$4d^{9}4f^{2}5p \rightarrow 4f^{2}$ transitions to the line formation should not 
be very large. We assume that the emission from some other tungsten ions forms 
lines in the experimental spectrum in this region. 

Another strong group of lines is formed by the $4f6g \rightarrow 4f^{2}$, 
$4d^{9}4f^{2}5f \rightarrow 4f^{2}$, and $4p^{5}4f^{2}5s \rightarrow 4f^{2}$ 
transitions. An unidentified line with the wavelength of $2.284\pm 0.002$ nm was 
observed in a tokamak spectrum \cite{2011cjp_89_591_podpaly}. The line was 
interpreted as the $4f6g \rightarrow 4f^{2}$ transitions in the CoBIT and LHD 
spectra \cite{2012aipcp_1438_91_sakaue, 2013aipcp_1545_143_morita}.  
Our calculations produce two peaks near $2.25$ nm for modeling intensities with 
the MCDF data and one peak at $2.260$ nm when the DW method is used in FAC to 
obtain the electron-impact excitation rates (Fig.~\ref{w26_corona_1_4}). The 
strongest lines correspond to the $4f6g \rightarrow 4f^{2}$ transitions. On the 
other hand, the $gA$ values of the MCDF data for the $4f6g \rightarrow 4f^{2}$ 
transitions form a peak at $2.285$ nm (Fig.~\ref{gA_1_4}a). All these 
wavelengths agree with the observation. 

Modeling predicts a peak at $1.996$ nm in the MCDF calculations corresponding to 
the $4f7g \rightarrow 4f^{2}$ transitions and at $1.993$ nm in the FAC 
calculations (Fig.~\ref{w26_corona_1_4}). The total $gA$ spectrum has a peak at 
$2.003$ nm (Fig.~\ref{gA_1_4}a). These theoretical wavelengths are in close 
agreement with the unidentified experimental line at $2.088\pm 0.002$ nm 
\cite{2011cjp_89_591_podpaly}. 

The experimental intensities of lines at $2.3$ nm are approximately two times 
weaker compared with the lines at $3.1$ nm, while our modeling with the DW 
rates gives a smaller value. It can be explained by the fact that the Maxwellian 
distribution for the electron velocities occurs in the fusion plasma, and the 
electron density is by a few orders of magnitude higher than in the EBIT plasma. 
However, the $gA$ values for $4f5g \rightarrow 4f^{2}$, 
$4f6g \rightarrow 4f^{2}$, and $4f7g \rightarrow 4f^{2}$ transitions have very 
similar relative magnitudes (Fig.~\ref{gA_1_4}).

\subsection{$4-7$ nm region}

\begin{figure}
 \includegraphics[scale=0.5]{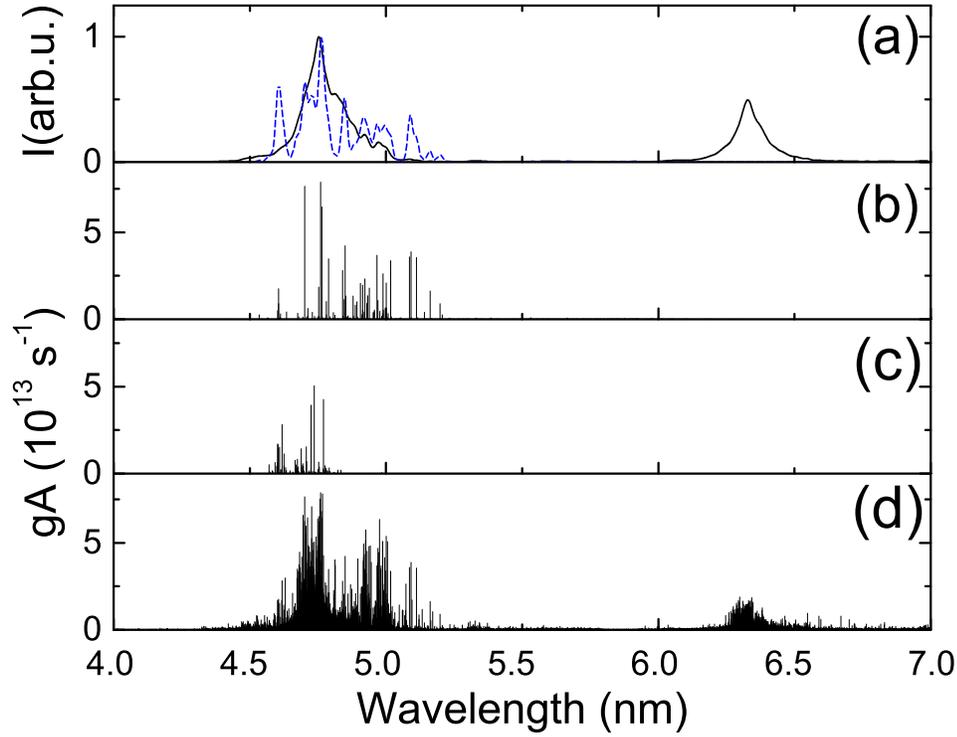}%
 \caption{
\label{gA_4_7} 
Calculated transition data for  W$^{26+}$ in the $4-7$ nm range;
(a) convoluted $gA$ spectra with a full width at half maximum  of 0.02 nm. 
The dashed line (blue online) corresponds to the sum of contributions from 
(b) $4d^{9}4f^{3} \rightarrow 4f^{2}$ and 
(c) $4f5d \rightarrow 4f^{2}$ transitions. The solid line corresponds to 
(d) the spectrum of all transitions considered in this work. }
\end{figure}

\begin{figure}
 \includegraphics[scale=0.5]{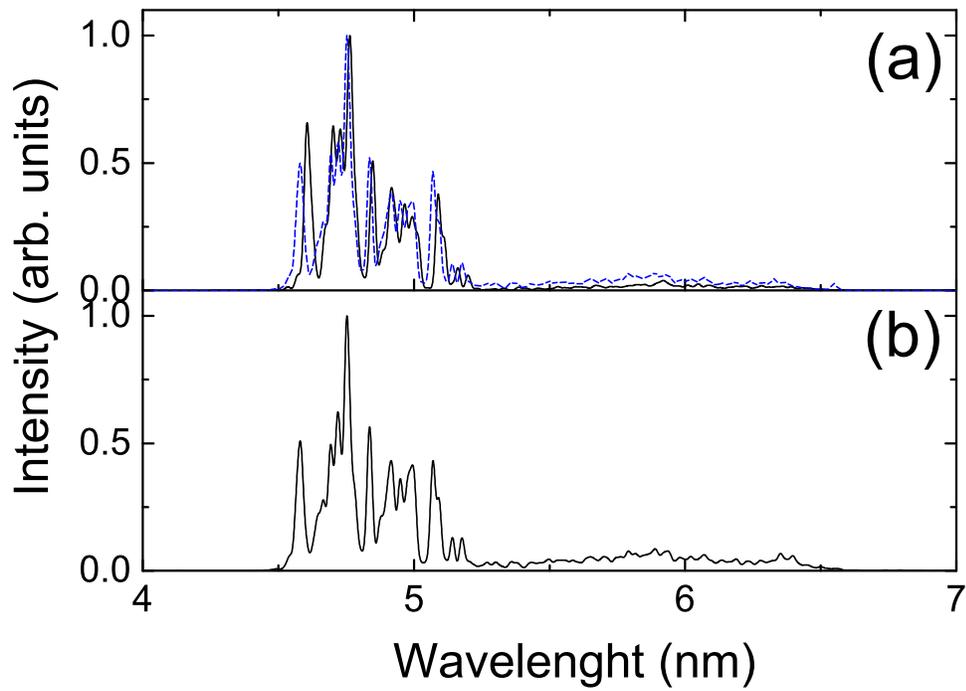}%
 \caption{
\label{w26_corona_4_7}
Same as Fig.~\ref{w26_corona_1_4}, but for the $4-7$ nm range. 
}
\end{figure}

Region of $4-7$ nm is covered by strong lines originating from the 
$4d^{9}4f^{3} \rightarrow 4f^{2}$ and $4f5d \rightarrow 4f^{2}$ transitions. 
The $gA$ spectra for these transitions are presented in Fig.~\ref{gA_4_7}. 
It can be seen that lines from the  $4f5d \rightarrow 4f^{2}$ transitions are 
concentrated on the shorter wavelength side while the lines from the 
$4d^{9}4f^{3} \rightarrow 4f^{2}$ transitions are spread more widely. Again, 
quite a good agreement is obtained among theoretical spectra 
(Fig.~\ref{w26_corona_4_7}). As mentioned above, the configuration mixing 
increases the radiative transition probabilities for the  
$4f5d \rightarrow 4f^{2}$ transitions by an order of magnitude 
(Fig.~\ref{ci_4f5d_4f2}).  However, the number of the 
$4d^{9}4f^{3} \rightarrow 4f^{2}$ transitions is much larger than the number 
of the $4f5d \rightarrow 4f^{2}$ transitions, and the former transitions 
dominate in  spectra. Another interesting feature in 
this region is an additional weaker structure of the lines with wavelengths 
from $5.5$ to $6.5$ nm. The fusion spectra of tungsten ions contain this group 
of lines \cite{2008ppcp_50_085016_Putterich, 2005jpb_38_3071_putterich}. 
However, the EBIT plasma does not have these lines 
\cite{2001pra_64_012720_radtke}. 


The collisional-radiative modeling using the HULLAC code \cite{BarShalomHULLAC} 
indicates that the strongest lines in the $5$ nm region are formed only by the 
$4d^{9}4f^{3} \rightarrow 4f^{2}$ transitions. Our results demonstrate some  
contribution from the $4f5d \rightarrow 4f^{2}$ transitions. Agreement for
wavelengths between our and the HULLAC calculations is within $0.04$ nm. 

Unfortunately, a strong emission of many ions in this wavelength region makes 
it impossible to identify lines in the fusion spectra. The previous studies 
\cite{2007pfr_2_s1060_chowdhuri, 2008ppcp_50_085016_Putterich, 
2011cjp_89_591_podpaly} established that some peaks in the emission band can be 
assigned to the transitions from ions of the higher ionization stages.


\newpage

\subsection{$10-30$ nm region}

\begin{figure}
 \includegraphics[scale=0.5]{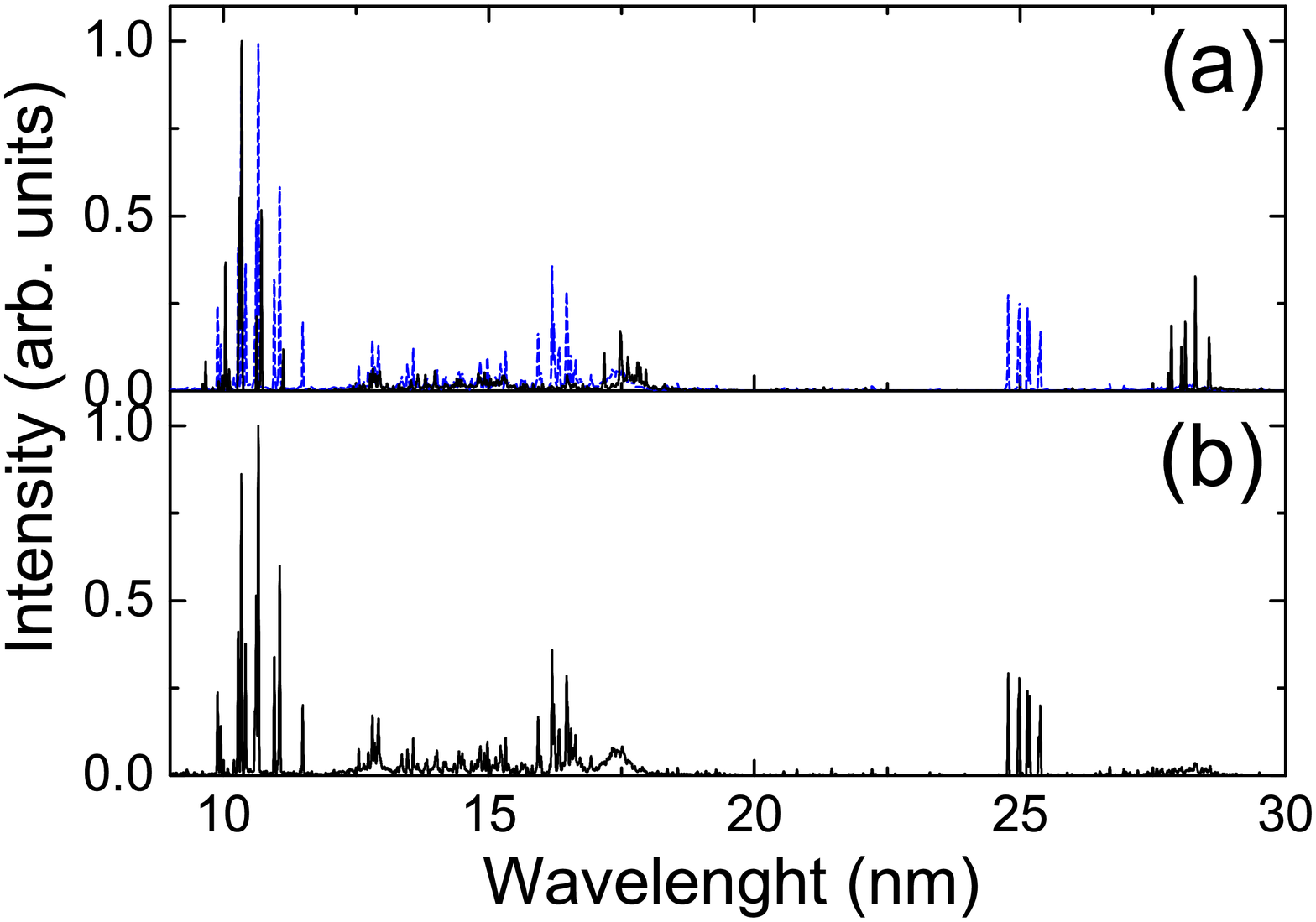}%
 \caption{
\label{w26_corona_10_30} 
Same as Fig.~\ref{w26_corona_1_4}, but for the $10-30$ nm range. 
}
\end{figure}

Lines in the range from 10 to 30 nm correspond mainly to the ($n=5$)$-$($n=5$) 
transitions in the modeled spectra (Fig.~\ref{w26_corona_10_30}). However, the 
strongest lines arise from the $4f5s \rightarrow 4f^{2}$ transitions, which 
concentrate in the range $10-11$ nm. As in the W$^{25+}$ ion case, the  
modeling predicts very small intensities of these transitions compared with the 
($n=5$)$-$($n=5$) transitions when the radiative cascade is not taken into 
account.


\begin{figure}
 \includegraphics[scale=0.6]{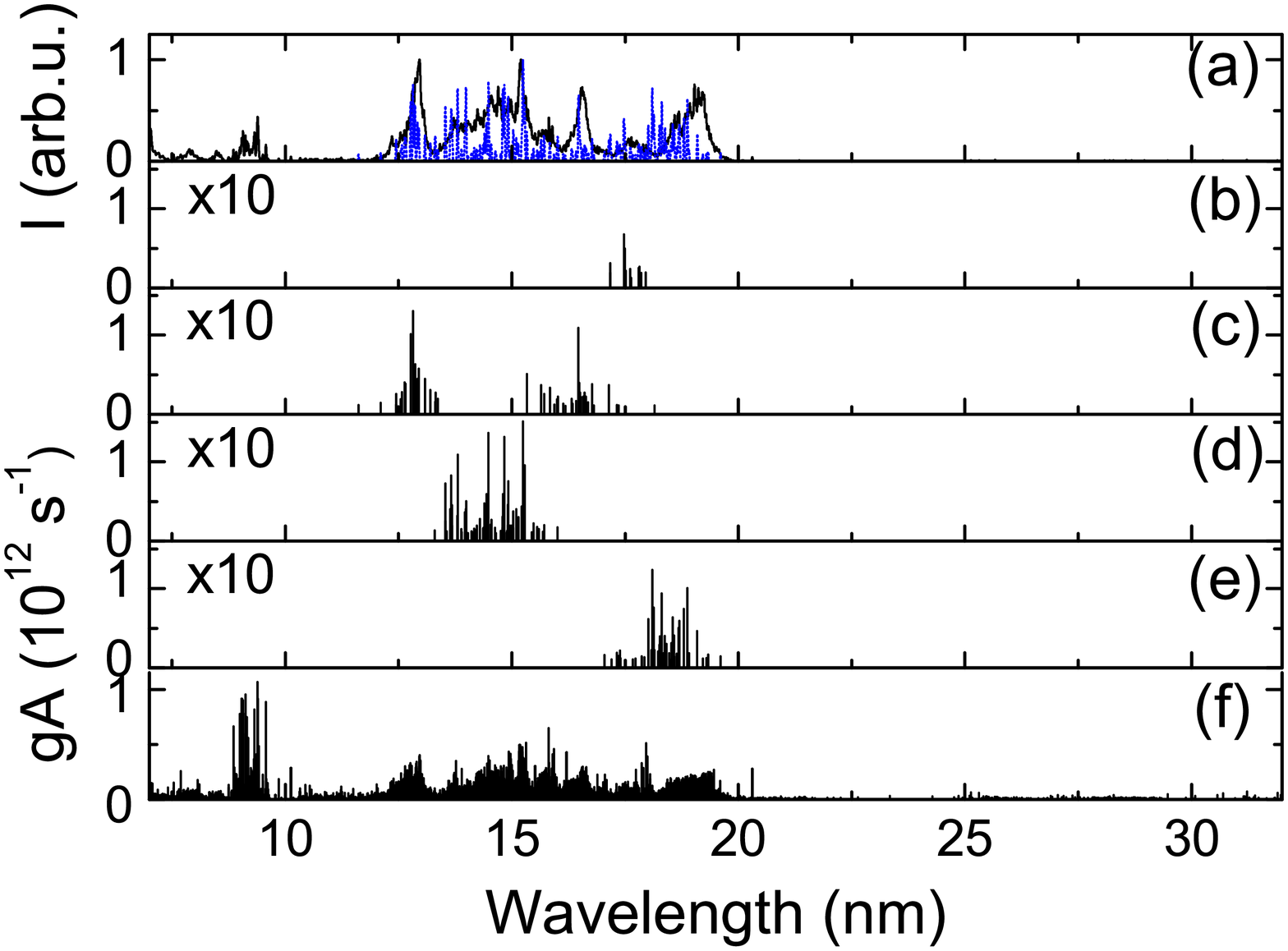}%
 \caption{
\label{gA_10_30}  
Calculated transition data for W$^{26+}$ in the $10-30$ nm range; 
(a) convoluted $gA$ spectra with a full width at half maximum  of 0.02 nm. 
The  dashed line (blue online) corresponds to the sum of contributions from 
(b) $4f5p \rightarrow 4f5s$, 
(c) $4f5d \rightarrow 4f5p$, 
(d) $4f5f \rightarrow 4f5d$, and 
(e) $4f5g \rightarrow 4f5f$ transitions. The solid line corresponds to 
(f) spectrum of all transitions considered in this work.  
}
\end{figure}


Another prominent group of lines is formed by the $4f5p \rightarrow 4f5s$ 
transitions at the long-wavelength side. The $4f5p$ and $4f5s$ configurations 
are the first two excited ones in the W$^{26+}$ ion. The radiative decay from 
these configurations to the ground one occurs through the forbidden transitions 
in the single-configuration approach. It can be seen from the energy level 
spectrum that two groups of the energy levels are formed by the $4f5p$ 
configuration (Fig. \ref{energy}). The higher structure corresponds to the states 
with the $5p_{3/2}$ subshell, while the lower structure is formed by the 
$5p_{1/2}$ subshell. Transitions from the higher-lying levels dominate in the 
$gA$ spectrum (Fig. \ref{gA_10_30}). The lines with the shorter wavelengths for 
the $4f5p \rightarrow 4f5s$ transitions have approximately 10 times larger 
$gA$ values compared to the longer wavelengths. However, the situation 
drastically changes in the corona and CRM spectra. The intensities of 
longer-wavelength transitions from the lower group of the energy levels 
strongly increase compared with the ones from the higher group. 

It has to be noted that, for the long-wavelength group of the
$4f5p \rightarrow 4f5s$ transitions, FAC produces wavelengths approximately 
$3$ nm smaller compared with the MCDF calculations. On the other hand, it has 
been previously determined that discrepancy between the theoretical wavelengths 
obtained with FAC and the experimental values can reach up to $2$ nm for some 
transitions in Er-like tungsten. Compared with the 
$4f^{2}5p \rightarrow 4f^{2}5s$ transitions in the W$^{25+}$ ion, the current 
wavelengths calculated with FAC are shorter by approximately $3$ nm. We consider 
that this difference of the wavelengths for two neighboring ions is too large. 
In our opinion, MCDF provides more accurate wavelengths. 

The other groups of lines between 12 to 18 nm are formed by the 
$4f5d \rightarrow 4f5p$ ($12-13$ nm), $4f5f \rightarrow 4f5d$ ($13-15$ nm), and 
$4f5p \rightarrow 4f5s$ ($17-18$ nm) transitions.

The current modeling demonstrates that this region is covered by lines that do 
not form a band structure and can be of interest for EBIT measurements.

\section{Conclusions}

Energy levels, radiative transition wavelengths and probabilities have been 
studied for the W$^{26+}$ ion. The multiconfiguration Dirac-Fock  method and 
the Dirac-Fock-Slater approach were employed to calculate atomic data using 
the relativistic configuration interaction method. Our calculations demonstrate 
that the radiative lifetimes of some levels from the $4d^{9}4f^{3}$ 
configuration have values exceeding the lifetimes of the levels from the ground 
configuration.

The configuration interaction strength has been used to determine influence of 
the correlation effects in the W$^{26+}$ ion. The mixing of configurations opens 
decay paths for the $4f5s$ configuration to the ground configuration through the 
electric dipole transitions. This mixing increases the total radiative decay 
rates from $10$ s$^{-1}$ to $10^{5}$ s$^{-1}$. A strong influence of the 
correlation effects is also determined for the $4f5d \rightarrow 4f^{2}$ 
transitions. The transition probabilities for these transitions increase by an 
order of magnitude when the configuration mixing is taken into account. 

The corona and collisional-radiative models are used to determine the influence 
of various transitions on the formation of lines in a low-density plasma. Two 
approaches are applied to calculate the line intensities for the corona model. 
In one of them, the electric dipole line strengths are used instead of the 
electron-impact excitation rates for the excitations from the levels of the 
ground configuration. In another approach, the DW method is employed to 
calculate the collisional excitation rates. The analysis of the strongest DW 
cross-sections for the excitations from the  levels of the ground configuration
reveals that the dominant part of the excitations corresponds to the 
$\Delta J = 0, \pm 1$ transitions. It is one of the reasons, in addition to the 
strong mixing of states and the radiative cascades from the excited 
configurations, why good agreement between both approaches for spectral line 
intensities occur. Good agreement between the corona and collisional-radiative 
spectra demonstrates that the simple approach can be successfully applied to 
predict the strongest lines in the EBIT plasma.

The current results of the spectra modeling show that some unidentified lines 
in the fusion plasma can belong to the transitions from the $4f5g$, $4f6g$, and 
$4f7g$ configurations to the ground one of the W$^{26+}$ ion. It was deduced 
that the strongest group of potentially observable W$^{26+}$ lines in $10-30$ 
nm region corresponds to the $4f5s \rightarrow 4f^{2}$ transitions. The 
intensities of these lines increase due to the radiative cascades from the 
higher-lying levels. Modeling demonstrates that this region can be of interest 
for the EBIT measurements because the structure of lines that do not form 
emission bands appears in the  spectra.

\section*{Akcnowledgement}

This research was funded by European Social Fund under the Global Grant Measure
(No.: VP1-3.1-\v{S}MM-07-K-02-015).

\end{document}